\documentclass{aa}  

\usepackage{booktabs}
\usepackage{dblfloatfix}
\usepackage{subcaption}
\usepackage{graphicx}
\usepackage{float} 
\usepackage{appendix}
\usepackage{txfonts}
\usepackage{hyperref}
\hypersetup{
    colorlinks=true,
    linkcolor=blue,
    filecolor=magenta,      
    urlcolor=cyan,
    citecolor=blue,
}
\let\textAA\AA
\renewcommand{\AA}{\ifmmode\text{\textAA}\else\textAA\fi}
\begin{document} 

   \title{Outflows in the early universe: neutral gas absorption in galaxies at $z>3$ from low-resolution JWST spectroscopy}


   \titlerunning{Neutral gas absorption in galaxies at $z>3$}

   \author{Matteo Sapori\inst{\ref{unibo}, \ref{inaf}} \and
   Sirio Belli\inst{\ref{unibo}} \and
   Letizia Bugiani\inst{\ref{unibo}, \ref{inaf}} \and
   Amir H. Khoram\inst{\ref{unibo}, \ref{inaf}} \and
   Caterina Liboni\inst{\ref{unibo}, \ref{inaf}} \and
   Gabriel Maheson\inst{\ref{unibo}} \and
   Gabriel Brammer\inst{\ref{copenhagen}} \and
   Rebecca Davies\inst{\ref{swinburne}} \and
   Pratika Dayal\inst{\ref{cita}, \ref{david}, \ref{Sngiorgio}} \and
   Anna de Graaff\inst{\ref{mpia}} \and
   Joel Leja\inst{\ref{psu1}, \ref{psu2}, \ref{psu3}} \and
   Rohan Naidu\inst{\ref{hawaii}} \and
   Pascal Oesch\inst{\ref{geneva}, \ref{copenhagen}} \and
   Stefano Sotira\inst{\ref{unibo}} \and
   Sandro Tacchella\inst{\ref{cambridge}, \ref{stedmunds}} \and
   Bingjie Wang\inst{\ref{princeton}} \and
   Katherine E. Whitaker\inst{\ref{umass}}}
   \institute{
   \label{unibo} Dipartimento di Fisica e Astronomia, Università di Bologna, Bologna, Italy \and
   \label{inaf} INAF Astrophysics and Space Science Observatory Bologna, Bologna, Italy \and
   \label{copenhagen} Cosmic Dawn Center (DAWN), Niels Bohr Institute, University of Copenhagen, Copenhagen, Denmark \and
   \label{swinburne} Centre for Astrophysics and Supercomputing, Swinburne University of Technology, Hawthorn, VIC, Australia \and
   \label{cita} Canadian Institute for Theoretical Astrophysics, University of Toronto, Toronto, ON, Canada \and
   \label{david} David A. Dunlap Department of Astronomy and Astrophysics, University of Toronto, 50 St George St, Toronto ON M5S 3H4, Canada \and
   \label{Sngiorgio} Department of Physics, 60 St George St, University of Toronto, Toronto, ON M5S 3H8, Canada \and
   \label{mpia} Max Planck Institute for Astronomy, Heidelberg, Germany \and
   \label{psu1} Department of Astronomy \& Astrophysics, The Pennsylvania State University, University Park, PA, USA \and
   \label{psu2} Institute for Computational \& Data Sciences, The Pennsylvania State University, University Park, PA, USA \and
   \label{psu3} Institute for Gravitation and the Cosmos, The Pennsylvania State University, University Park, PA, USA \and
   \label{hawaii} Institute for Astronomy, University of Hawai'i, Honolulu, HI, USA \and
   \label{geneva} Department of Astronomy, University of Geneva, Geneva, Switzerland \and
   \label{cambridge} Kavli Institute for Cosmology, Department of Physics, University of Cambridge, Cambridge, UK \and
   \label{stedmunds} St Edmund's College, University of Cambridge, Cambridge, UK \and
   \label{princeton} Department of Astrophysical Sciences, Princeton University, Princeton, NJ, USA (NHFP Hubble Fellow) \and
   \label{umass} Department of Astronomy, University of Massachusetts, Amherst, MA, USA 
   }

   \date{Received ???; accepted ???}

 
\abstract{
Recent JWST/NIRSpec observations have shown that Na~I~D absorption tracing neutral gas outflows is widespread in massive galaxies at Cosmic Noon ($z\sim2$--$3$), but their prevalence at higher redshift remains largely unexplored. Here we investigate whether similar outflows are already in place during the first 2\,Gyr of cosmic history ($z>3$), using a sample of 811 galaxies at $3<z<7$ drawn from the public DAWN JWST Archive and the Mirage or Miracle survey, selected to have secure spectroscopic redshifts and continuum SNR$>5$ in their JWST/NIRSpec prism low-resolution spectra ($R\sim100$). We derive the physical properties and star formation histories of these galaxies via \texttt{prospector} SED fitting and isolate the Na~I~D feature in each spectrum by subtracting the best-fit stellar continuum. We detect an excess of Na~I~D absorption in 20 galaxies, almost all at $3<z<5$ with stellar mass $>10^{10}\,M_\odot$, corresponding to an overall detection fraction of $\sim2.5\%$. The detection fraction rises steeply with stellar mass and quiescence, reaching $\sim11\%$ among the most massive galaxies and $\sim43\%$ among massive quenched systems. The Na~I~D equivalent widths are large, spanning the range from $4$ to $16\,\AA$, reaching the highest values in dusty star-forming galaxies. In previous medium-resolution observations, values of EW above $5\,\AA$ were mostly found in outflowing gas; we thus interpret our detections as mostly tracing neutral gas outflows despite our inability to measure the gas kinematics from the prism low-resolution spectra. We also perform a stacking analysis, which confirms the trends with physical properties and provides average values of Na I D EW for different subsamples.
Under conservative assumptions, we estimate mass outflow rates of $4$ to $12\,M_\odot\,\mathrm{yr}^{-1}$, exceeding the current star formation rate for about 30\% of the detected galaxies. We conclude that neutral outflows are likely already present and important in massive galaxies during the first 2\,Gyr of cosmic history.}



   \maketitle

\section{Introduction}
Outflows are a fundamental component in the framework of galaxy evolution, acting as crucial regulators of the star formation history (SFH) and the chemical enrichment of the interstellar medium (ISM). These flows of gas are driven by stellar processes or by active galactic nuclei (AGN) (\citealt{Rupke_2018};\citealt{2020A&ARv..28....2V}) and may play a key role in galaxy quenching (e.g. \citealt{2017MNRAS.465.3291W}; \citealt{Man&Belli+18}).
However, obtaining a complete census of galaxy outflows is complicated due to their multi-phase nature, since we can typically study only one gas phase at the time. While ionized outflows are readily accessible through strong emission lines, even at high redshift (e.g. \citealt{2019ApJ...875...21F}; \citealt{2016MNRAS.459.3144Z}, \citealt{2024A&A...685A..99C}), we know that their density, and consequently the mass they carry, are often insufficient to strongly impact the host galaxy \citep{Concass+22}. Indeed, theoretical predictions suggest that the bulk of the gas mass is expected to reside in the neutral atomic or molecular phases (\citealt{Hopkins+20}; \citealt{Kim+20}) where gas ejection is more efficient \citep{fiore_agn_2017}. Historically, observing these colder components has been challenging (\citealt{2020A&ARv..28....2V}), particularly in low-metallicity or distant galaxies, leading to an incomplete understanding of the total mass, momentum, and energy carried away by outflows.
The advent of the James Webb Space Telescope (JWST) is now revolutionizing our understanding of galaxy outflows in the early universe. Its unprecedented sensitivity and high-resolution infrared spectroscopy grant access to the key rest-frame optical diagnostics at Cosmic Noon (z $\sim$ 2-3) and beyond. Specifically, JWST enables the use of the Na I D resonance doublet as a robust tracer of neutral atomic gas for the first time at high redshift.
The Na I D doublet is an optimal choice for identifying and characterizing cold outflows: since its first ionization potential (5.1 eV) is significantly lower than that of hydrogen (13.6 eV), Na I D exists only in neutral, cold, and well-shielded regions where gas and dust are present (e.g. \citealt{Baron_2020}).
Recent studies using JWST/NIRSpec have confirmed that neutral-phase outflows are common at Cosmic Noon, in agreement with earlier rest-frame UV absorption studies (e.g. \citealt{davies_jwst_2024}; \citealt{Liboni+25}); crucially, Na I D also enables a direct measurement of the mass carried by the outflow. Indeed, these new observations show that the mass carried by the neutral component of the outflow can be more than one order of magnitude larger than that carried by the ionized component \citep{belli_star_2024, D_Eugenio+24, Sun+26}, similarly to what is found in the local universe (\citealt{Avery+22}; \citealt{Robert-Borsani+20}).

There have been many detections of strong neutral outflows in galaxies with minimal ongoing star formation, and their derived mass-loading factors substantially exceed unity, suggesting that these outflows could be powerful enough to suppress star formation and drive rapid quenching \citep{belli_star_2024, D_Eugenio+24}. However, the link between outflow activity and quenching is not straightforward: outflow return timescales estimated for quiescent galaxies at $z\sim3.5$ are $\sim3$--$180$\,Myr, much shorter than the time elapsed since the last major starburst, as inferred from the stellar population ages \citep{zhu2026againneutraloutflowsz35}. This mismatch indicates that the outflows currently observed are unlikely to be relics of the star-forming episode that assembled most of the stellar mass, and may instead reflect short-lived fountain cycles rather than ejective quenching events. Most studies attribute the powerful outflows seen in quiescent galaxies to AGN feedback: current star formation is energetically insufficient to power them \citep{davies_jwst_2024, Sun+26sample}, and yet many of these systems are not classified as AGN on standard optical diagnostics (e.g., the BPT diagram) or via X-ray detection, possibly because they host intrinsically low-luminosity AGN \citep[e.g.,][]{Bugiani+25}, or because the outflows are fossils of past AGN episodes \citep{Sun+26, Taylor+26}.

While it is now well established that outflows are common in massive galaxies at $z \sim 2$, their impact remains highly uncertain at higher redshift ($z>3$). This is the critical epoch when the first massive quiescent galaxies appear \citep{carnall_2023, Weibel_Graaf+25}; feedback must be extremely efficient at this redshift to regulate the swift growth of galaxies, which is fueled by high gas fractions. Specifically, massive outflows in these young systems are paramount because they represent the only known mechanism capable of rapidly transporting metals out of the galaxy and into the circumgalactic medium (CGM) and beyond, seeding the enrichment of the CGM and intergalactic medium observed at later cosmic times.
JWST observations have led to the discovery of neutral outflows at $z\sim3-7$ in small samples of star-forming \citep{Davies+26} and quiescent galaxies \citep{Po-Feng+25,Valentino+25,zhu2026againneutraloutflowsz35, Taylor+26}, confirming that outflows were already in place in the early universe.

In this work, we carry out a systematic study of neutral outflows detected by Na~I~D absorption in the galaxy population at $3<z<7$, using archival NIRSpec data obtained with the low-resolution prism. Our approach is complementary to the JWST studies of neutral outflows cited above, which employ medium-resolution observations with the NIRSpec gratings. By using prism spectra we lose in velocity resolution but gain in sample size; we can thus obtain a comprehensive view of neutral gas in galaxies at high redshift, but we are limited to the study of only the strongest outflows.

The layout of this paper is as follows. In Section~\ref{sec: data}, we describe the observational data and the construction of the galaxy sample. In Section~\ref{sec: Methods}, we present the methodology adopted for the SED fitting and the extraction of the Na~I~D absorption signal. In Section~\ref{sec: results} we investigate the relation between Na~I~D absorption and the physical properties of the host galaxies. We then discuss the interpretation of the Na~I~D excess as a tracer of neutral outflows, and derive the outflow properties, in Section~\ref{sec: outflows}. Finally, we summarize our main conclusions in Section \ref{sec: conclusions}.

Throughout this paper, we assume a \citet{Chabrier_IMF} initial mass function (IMF) and a flat $\Lambda$CDM cosmology with [$\Omega_m$, $\Omega_\Lambda$, h] = [0.3, 0.7, 0.7].

\begin{figure*}[tbph] 
\centering
    \includegraphics[scale=0.48]{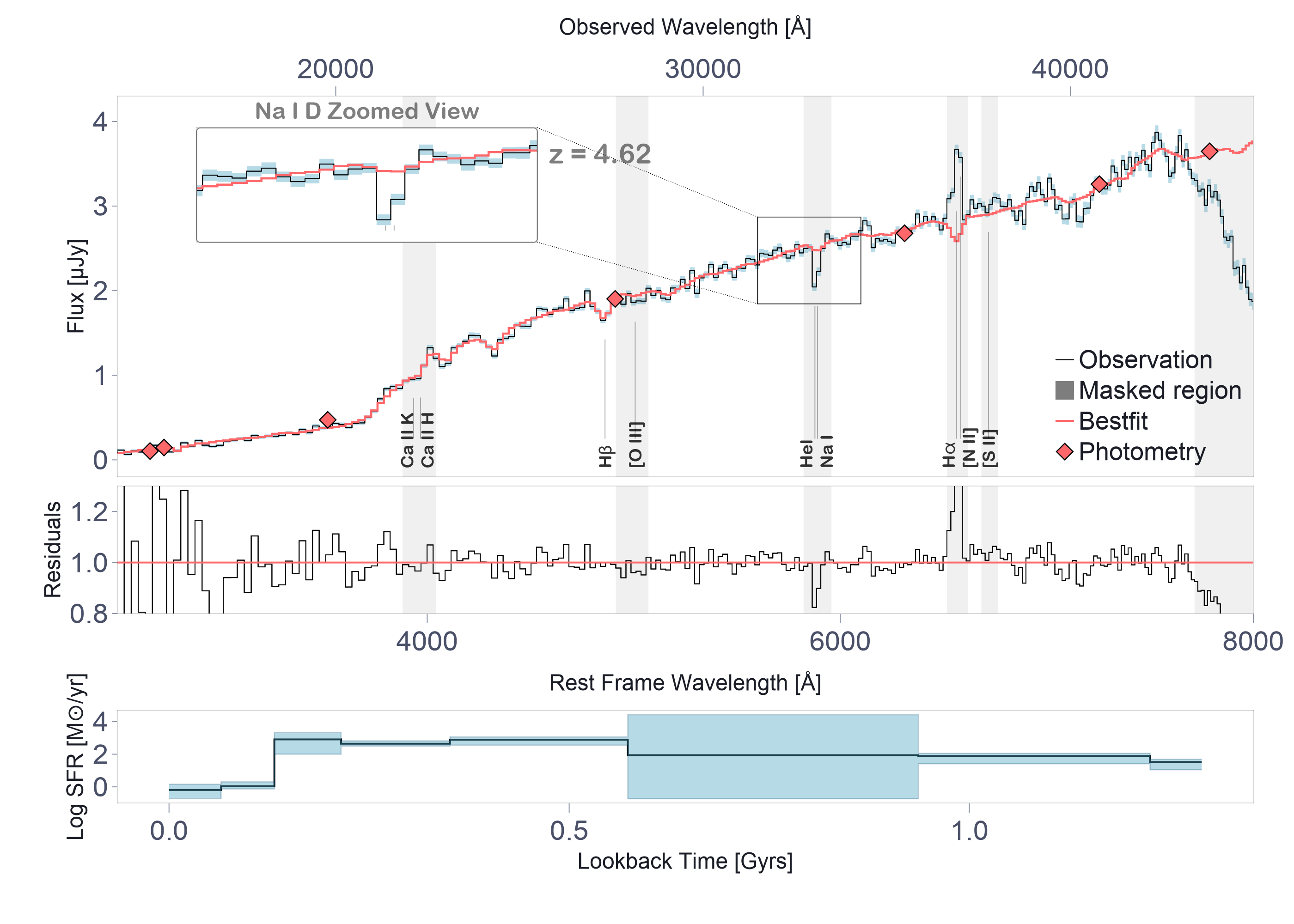}
    \caption{Best-fit SED and derived SFH. \textbf{Top panel}: The observed NIRSpec prism spectrum (blue) and photometric data (red diamonds) are shown against the best-fit model from \texttt{prospector} (red line) for the source \text{rubies-uds23-v4-prism-clear-4233-140707}. The best-fit model shown is regenerated without emission lines, to display the underlying stellar continuum. The data span from the rest-frame UV to the near-infrared. Significant absorption lines (e.g., Ca II, Na I D) and emission lines (e.g., [O III], H$\alpha$) are marked. The inset shows a detailed view of the Na I D feature. \textbf{Middle panel}: The residuals of the fit (Observation/Bestfit). \textbf{Bottom panel}: The non-parametric SFH reconstructed from the fit. The solid line shows the median SFR at each time bin, and the shaded region marks the 68\% confidence interval.}
    \label{fig: fit example}
\end{figure*}

\section{Data}
\label{sec: data}

Conducting a systematic analysis of galaxy outflows at high redshift requires a large homogeneous sample, which we draw from the DAWN JWST Archive (DJA\footnote{\url{https://dawn-cph.github.io/dja/index.html}}). The DJA is a publicly accessible and non-proprietary repository for reduced JWST galaxy observations; all data products within the archive are processed using \texttt{grizli} \citep{brammer_grizli} and \texttt{msaexp} \citep{brammer_msaexp} as described in \citealt{Graaf_rubies+25}.
We decide to use spectra obtained with the NIRSpec prism, which offer two key advantages. First, the DJA contains a substantial number of prism observations, enabling a statistically meaningful analysis. Second, the wide wavelength coverage together with the depth provided by the prism allow us to characterize the stellar continuum with high fidelity, and to constrain important physical properties of the galaxies, such as their mass and star formation rate.
Adopting version V4.2 of the DJA, we select all sources with spectroscopic redshift in the range $3 < z < 7$ and with a quality flag (grade) of 3, indicating a high-confidence redshift measurement. The upper bound of $z<7$ is set by the requirement of having the Na I D resonant absorption line (our key diagnostic feature) within the spectral coverage of the NIRSpec prism.
We further require a continuum SNR$ > 5$ (average over the whole spectrum) to ensure a data quality sufficient for the detection of Na~I~D absorption. This initial filter yields a sample of 978 sources from the DJA.
In addition to our spectroscopic data we utilize complementary photometry from ASTRODEEP-JWST \footnote{\url{http://www.astrodeep.eu/}}. The ASTRODEEP photometric catalogs \citep{Merlin_2024} incorporate 16 bands, including new HST reductions, spanning from 0.44$\mu$m to 4.44$\mu$m for approximately 530,000 detected sources. This public catalog is specifically designed to facilitate the study of high-redshift galaxy populations by consolidating data from eight key JWST NIRCam programs: Abell 2744 (including GLASS-JWST \citealt{glass_survey}, UNCOVER \citealt{uncover_survey}), EGS (CEERS; \citealt{CEERS_Survey}), COSMOS and UDS (PRIMER; \citealt{primer_survey}), and GOODS North and South (JADES \citealt{jades_survey} and NGDEEP \citealt{ngdeep_survey}). Our targets are cross-matched against the ASTRODEEP catalog using a 1 arcsecond search radius to associate a photometric source with each spectrum. We find a match for 778 sources.
Finally, we visually inspect the sample to discard spectra that are not galaxies, where data reduction failed catastrophically, or that contain gaps or invalid pixels within or nearby the Na I D region, removing 10 sources from the DJA sample.
To this DJA sample of 768 sources, we added 43 galaxies from the Mirage or Miracle \citep[MoM,][]{mirage_or_miracle} survey, selected applying the same quality criteria described above (spectroscopic redshift in the range $3<z<7$, grade 3, continuum SNR$ > 5$, and visual inspection).
The entire process yields a final comprehensive sample consisting of 811 sources.

\section{Methods}
\label{sec: Methods}
\subsection{Prospector SED Fitting} 
\label{sec: sed fitting}
In this section, we describe the methodology employed for the Spectral Energy Distribution (SED) fitting throughout this work.
To infer the physical properties of our galaxies we utilize \texttt{prospector} \citep{2021ApJS..254...22J}, a Bayesian stellar population inference code, to perform a joint fit of both the NIRSpec prism spectroscopy and the photometry measured from JWST and HST (see Section~\ref{sec: data}).
For the modeling of the SEDs, \texttt{prospector} relies on the Flexible Stellar Population Synthesis (FSPS) library \citep{2010ascl.soft10043C}.
Our stellar population model is built upon the following configuration:
\begin{itemize}
\item We adopt the MIST isochrones \citep{2016ApJS..222....8D} and the MILES spectral libraries \citep{miles_sp_lib}.

\item The redshift is centered on the value inferred from the NIRSpec prism spectrum in the DJA archive with a tight Gaussian prior (sigma = 0.05), allowing a small flexibility in redshift in order to account for possible uncertainties.

\item We assume a Chabrier  Initial Mass Function \citep{Chabrier_IMF}.

\item For the SFH, we adopt a non-parametric model consisting of eight temporal bins, logarithmically spaced up to $0.9\,t_{\rm U}$, where $t_{\rm U}$ is the age of the Universe at the redshift of the source. This is implemented using the \textit{Continuity SFH} prior \citep{2019ApJ...876....3L} available in \texttt{prospector}, which penalizes large variations between adjacent time bins, thereby favoring smoother SFHs.

\item The dust attenuation is modeled using a two-component prescription which distinguishes between the attenuation from dust in the stellar birth clouds which applies to only stars younger than 10 Myr, and that from the diffuse interstellar medium which attenuates all stellar light, following the \citealt{charlot_2000_model} model.

\item We model the infrared dust emission following the model from \citealt{Draine+07}, which assumes energy conservation, such that the total energy absorbed by dust from the stellar radiation field is re-emitted entirely in the infrared.

\item We adopt a 10th-order polynomial to correct the NIRSpec spectrum for slit losses or other calibration systematics.

\item To account for underestimated uncertainties in the spectrum, we incorporate a ``jitter'' parameter that uniformly inflates the error bars improving the overall goodness-of-fit.

\item We fit only the total metallicity, leaving the relative abundances fixed to the solar pattern.

\item We include the emission lines in the model, following \citet{Byler+17}.

\item AGN models are not included.
\end{itemize}

This model configuration results in a total of 22 free parameters. We utilize the dynamic nested sampling library \texttt{dynesty} \citep{dynesty.493.3132S} to explore the parameter space and derive posterior probability distributions for each parameter.
For the analysis presented in this work, the ``best-fit'' value for a given parameter is taken as the maximum a posteriori (MAP) of its posterior distribution, and the $1\sigma$ uncertainties are defined by the 16th and 84th percentiles of the distribution.
Figure~\ref{fig: fit example} illustrates an example from our sample, with the best-fit model reproducing both the photometric points and the prism spectrum; the corresponding SFH is shown in the bottom panel.

\begin{figure*}[p]
\centering
    \includegraphics[scale=0.65]{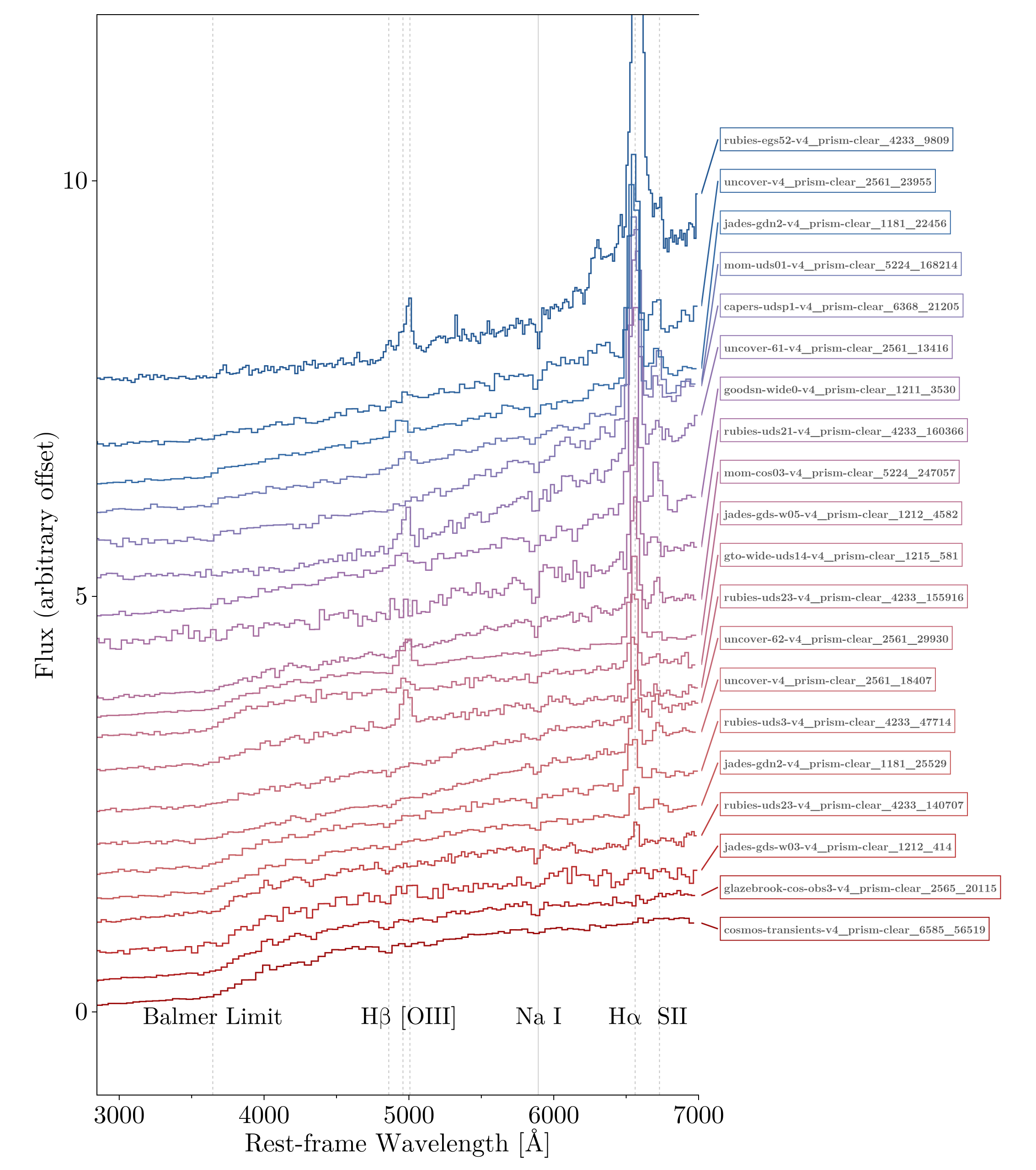}
    
    \caption{Rest-frame NIRSpec prism spectra of the of the 20 galaxies of the golden sample with robust Na I D detection. Spectra are normalized to the \texttt{prospector} best-fit flux in the rest-frame  $3000$--$7000$\,\AA\ window and offset vertically for clarity.
    Vertical dashed lines mark the position of important spectral features. Spectra are sorted and color-coded by H$\alpha$ flux for visualization purposes only, from dark red (H$\alpha$-faint) at the bottom to dark blue (H$\alpha$-bright) at the top. Individual galaxy identifiers are indicated on the right.}
    \label{fig: golden sample specs}
\end{figure*}

\begin{figure*}[ht] 
\centering
    \includegraphics[width=\linewidth]{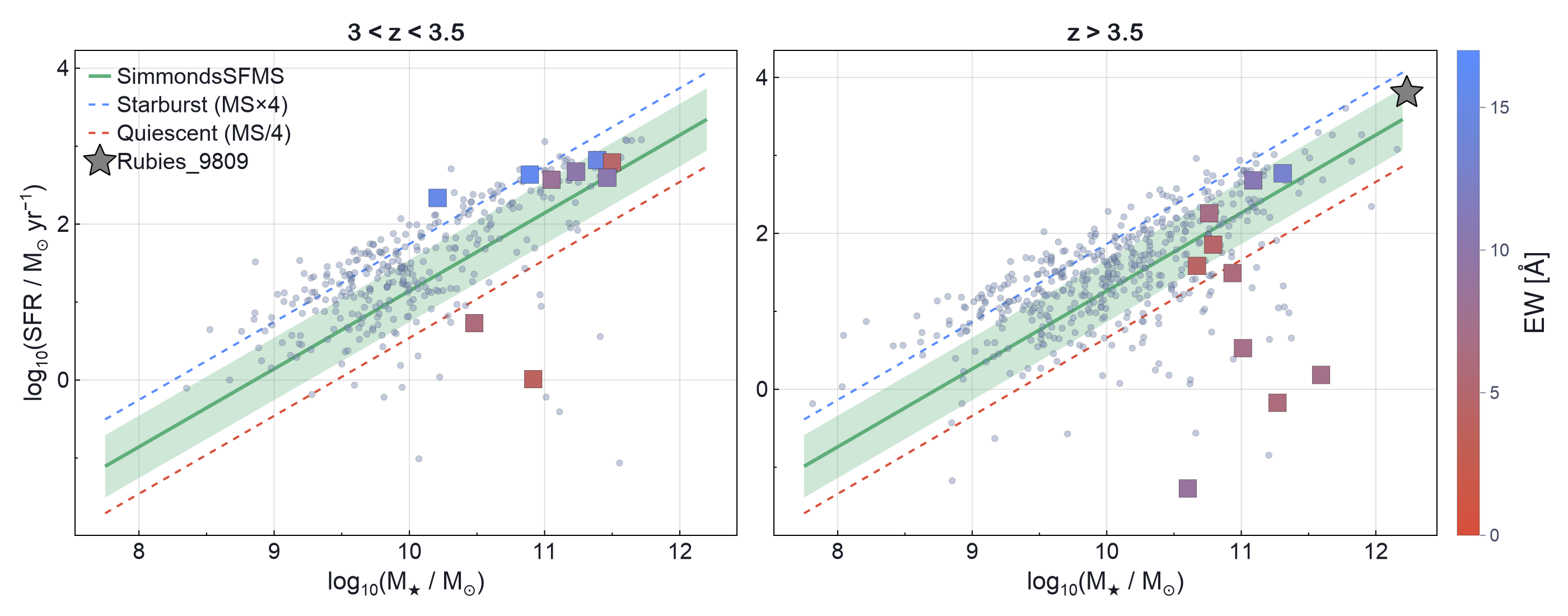}
    \caption{The distribution of the parent sample on the $M_*-SFR$ plane, divided into two redshift bins ($z=3$ to $z>3.5$). In each panel, the solid green line indicates the SFMS relation from \citealt{Simmonds_sfms+25}. The blue and red dashed lines mark our classification thresholds for 'starburst' ($SFR > 4 \times SFR_{\text{MS}}$) and 'quenched' ($\mathrm{SFR} < 0.25 \times SFR_{\text{MS}}$) galaxies, respectively. The full sample of sources is shown as light-blue circles. The golden sample (i.e., with robust Na~I~D detection) is shown with square markers, with the color indicating the measured EW of Na I. In the panel at $z>3.5$, the broad-line AGN (Rubies 9809) is plotted with a gray star; we caution that the reported SFR and stellar mass values are not well constrained.}
    \label{fig: sfms}
\end{figure*}

\subsection{Spectral Line Masking} 
\label{sec: line masking}
The Na I D absorption originates in both stars and cold neutral gas. In order to study the gas component, it is thus necessary to remove the stellar contribution from the spectra.
To ensure that our SED fitting is primarily constrained by the stellar continuum and to minimize the influence of nebular emission, we mask several prominent spectral lines prior to fitting with \texttt{prospector}.
Crucially, nebular emission is still included in the model to maintain self-consistency, since the contribution of strong emission lines is present in the broadband photometry.
This allows the model to derive robust stellar population properties such as stellar mass and SFH.  For consistency across our sample of 811 galaxies we apply a uniform masking scheme to all spectra, where we exclude strong nebular emission lines (e.g., [O III] doublet, H$\alpha$, and [S III]) and resonant absorption features (Na I D and Ca II H,K).
For each line, the masked region is defined as a fixed-width window of $5\sigma$, centered on the line wavelength, where $\sigma$ is computed from the Full Width at Half Maximum (FWHM) of the NIRSpec prism instrumental resolution at the corresponding wavelength.
Masked spectral regions are marked in gray in Figure~\ref{fig: fit example}.
To evaluate the impact of the masking procedure, we also perform an independent set of fits on the same spectra without excluding any spectral features. A direct comparison of the derived stellar population parameters reveals no significant systematic offsets between the masked and unmasked results.

\subsection{Na I D Equivalent Width Measurement}\label{sec: Na EW measurement}

We calculate the spectral residuals by dividing the observed data by the \texttt{prospector} best-fit model of the stellar continuum (shown in red in the top panel of Figure~\ref{fig: fit example}). The residual spectrum therefore retains the features that the stellar model does not capture, such as emission and absorption due to gas. If Na~I~D absorption by neutral gas is present, it will appear as an absorption line in the residual, as in the example shown in Figure~\ref{fig: fit example}. Stellar spectra also contain Na~I~D absorption, but this is included in the \texttt{prospector} best-fit model, and therefore does not contribute to the residuals.

In order to characterize the neutral gas properties, we thus investigate the Na~I~D absorption that appears in the residuals. 
The NIRSpec prism has a relatively low spectral resolution ($R = \lambda/\Delta\lambda \approx 30$--300), which means that the Na I D doublet ($\lambda\lambda$5890, 5896) is not resolved, preventing a reliable fit to the individual line profiles. We therefore measure the rest-frame equivalent width (EW) by integrating the residual spectrum over the full wavelength window masked during the SED fitting stage. 
Although the limited spectral resolution blends the doublet into a single feature, the EW is conserved under spectral convolution and is therefore not significantly affected by the instrumental resolution. However, the measurement is contaminated by blending with nearby spectral features, in particular the He I $\lambda$5877 emission line, which may originate from H II regions or AGN \citep[e.g.,][]{davies_jwst_2024}. In addition, resonant scattering can introduce emission that partially fills in the absorption trough \citep{Baron_2020}. For these reasons, the derived EW values should be regarded as lower limits to the intrinsic absorption strength.

\subsection{Sample with robust detections}
\label{sec: golden sample}
The NIRSpec data reduction pipeline is known to underestimate the flux uncertainties \citep[e.g.,][]{Maseda+23}, which could cause spurious detections of the Na~I~D absorption line. To ensure that the detections are robust, we rescale the error spectrum by introducing a multiplicative jitter factor estimated from two continuum regions adjacent to the Na I D window. In these sidebands we assume the continuum model to be accurate, such that the reduced chi-square between model and observation should satisfy $\chi^2_\nu = 1$. 
We compute the reduced chi-square in both continuum windows, average the two values, and interpret this mean as an estimate of the excess variance. The uncertainties in the Na I D region are then rescaled by the factor $\sqrt{\langle \chi^2_\nu \rangle}$, ensuring that the adopted errors reflect the level of fluctuations observed in the continuum. Across the sample, the inferred jitter parameter has a mean value of 1.54. This procedure is similar to what is done by the jitter parameter in \texttt{prospector}, but is performed on a small spectral region around Na~I~D.

For each galaxy, we measure the EW and its uncertainty using the rescaled error spectrum, and define their ratio as the signal-to-noise ratio $\mathrm{S/N}$ of the Na~I~D detection. Finally, we visually inspect each of these candidates in order to reject false positives, i.e. cases in which a large EW is produced by data reduction artifacts rather than by a genuine absorption feature.
We then define a ``golden sample'' comprising only the highest-confidence detections, with a signal-to-noise ratio $\mathrm{S/N} > 3$. 
The golden sample consists of 20 galaxies; their spectra are shown in Figure~\ref{fig: golden sample specs} and their main physical properties are summarized in Table \ref{tab:galaxy_properties}. The spectra with Na~I~D detection are from six surveys: JADES \citep{jades_survey}, the NIRSpec Wide GTO Survey \citep{Maseda+24}, RUBIES \citep{Graaf_rubies+25}, MoM \citep{mirage_or_miracle}, CAPERS (GO-6368, PI M. Dickinson), and UNCOVER \citep{uncover_survey};  and two more focused programs: a study of quiescent galaxies (GO 2565, PI: K. Glazebrook; \citealt{Nanayakkara+25}) and a study of high-redshift transients (DD 6585, PI: D. Coulter).

We note that the UNCOVER survey targets the lensing cluster field of Abell~2744 and is therefore subject to gravitational magnification. For the four galaxies in this field, we correct the \texttt{prospector}-derived stellar mass for the magnification factor reported in the UNCOVER lensing catalog. These corrected masses are the values reported in Table~\ref{tab:galaxy_properties} and used throughout the analysis. With the same procedure we also correct the stellar masses of all the UNCOVER galaxies in the parent sample.

None of the 20 galaxies in the golden sample, except for one case, shows broad H$\alpha$ emission, as measured by fitting a Gaussian profile convolved with the prism line-spread function to the H$\alpha$ line. This indicates that the emission of these galaxies is not dominated by AGN, and supports our choice not to include an AGN component in the \texttt{prospector} model. The only exception is rubies-egs52-v4\_prism-clear\_4233\_9809, a source at $z=5.70$ which has recently been classified as a Little Red Dot in some studies \citep{Setton+25}, but not in others \citep{Hviding+25, de_graaff_lrd+25}. This source is detected in the X-rays and at sub-mm wavelengths, which is exceptional at $z>5$, and suggests the presence of a broad-line AGN together with a dusty starburst. Because of its peculiar nature, this galaxy is marked with a distinct symbol throughout our figures, and its stellar mass as derived by \texttt{prospector} is highly uncertain.

We also verify the incidence of strong AGN emission in the parent sample using emission line ratios measured from archival NIRspec spectra at higher resolution, when available. The results, presented in Appendix~\ref{sec: app-agn}, show that the majority of the sample with available data do not present evidence for strong AGN emission.

\begin{table*}
\centering
\caption{Physical properties of the galaxy sample with robust detection of Na~I~D}
\label{tab:galaxy_properties}
\begin{tabular}{lrrrrrr}
\toprule
Galaxy & $\log M_\star$ & $\log \mathrm{SFR}$ & $z_{\mathrm{fit}}$ & S/N50 & $A_V$ & EW Na~I~D \\
\midrule
capers-udsp1-v4\_prism-clear\_6368\_21205 & 10.21 & 2.34 & 3.23 & 29.0 & 2.35 & $15.2 \pm 2.3$ \\
cosmos-transients-v4\_prism-clear\_6585\_56519 & 10.92 & 0.01 & 3.04 & 43.6 & 0.69 & $3.9 \pm 0.9$ \\
glazebrook-cos-obs3-v4\_prism-clear\_2565\_20115$^{\textcolor{blue}{a}}$ & 11.59 & 0.18 & 3.73 & 39.6 & 0.28 & $7.2 \pm 0.7$ \\
goodsn-wide0-v4\_prism-clear\_1211\_3530 & 11.39 & 2.82 & 3.30 & 28.0 & 1.05 & $14.8 \pm 1.5$ \\
gto-wide-uds14-v4\_prism-clear\_1215\_581 & 10.76 & 2.26 & 3.85 & 21.2 & 0.50 & $6.9 \pm 2.0$ \\
jades-gdn2-v4\_prism-clear\_1181\_22456 & 11.23 & 2.67 & 3.12 & 54.8 & 1.70 & $10.1 \pm 0.8$ \\
jades-gdn2-v4\_prism-clear\_1181\_25529 & 10.48 & 0.73 & 3.12 & 37.5 & 0.62 & $6.0 \pm 0.8$ \\
jades-gds-w03-v4\_prism-clear\_1212\_414 & 10.60 & -1.27 & 3.79 & 8.8 & 0.29 & $8.9 \pm 1.4$ \\
jades-gds-w05-v4\_prism-clear\_1212\_4582 & 11.50 & 2.79 & 3.21 & 65.8 & 0.59 & $4.7 \pm 0.8$ \\
rubies-egs52-v4\_prism-clear\_4233\_9809$^{*,\textcolor{blue}{b}}$ & 12.23 & 3.80 & 5.70 & 8.2 & 3.68 & $16.3 \pm 1.5$ \\
rubies-uds21-v4\_prism-clear\_4233\_160366$^{\textcolor{blue}{c}}$ & 11.31 & 2.77 & 3.93 & 10.4 & 1.64 & $13.1 \pm 1.9$ \\
rubies-uds23-v4\_prism-clear\_4233\_140707$^{\textcolor{blue}{c}}$ & 11.27 & -0.17 & 4.63 & 20.5 & 0.59 & $6.1 \pm 0.7$ \\
rubies-uds23-v4\_prism-clear\_4233\_155916$^{\textcolor{blue}{c}}$ & 11.01 & 0.53 & 4.10 & 15.9 & 0.40 & $7.1 \pm 1.4$ \\
rubies-uds3-v4\_prism-clear\_4233\_47714$^{\textcolor{blue}{c}}$ & 11.05 & 2.57 & 3.23 & 27.0 & 1.60 & $8.5 \pm 2.1$ \\
uncover-61-v4\_prism-clear\_2561\_13416$^{\textcolor{blue}{L}}$ & 11.09 & 2.68 & 4.03 & 24.0 & 2.18 & $10.8 \pm 0.4$ \\
uncover-62-v4\_prism-clear\_2561\_29930$^{\textcolor{blue}{L}}$ & 10.79 & 1.86 & 3.98 & 30.6 & 1.06 & $4.8 \pm 0.8$ \\
uncover-v4\_prism-clear\_2561\_18407$^{\textcolor{blue}{L}}$ & 10.67 & 1.58 & 3.98 & 43.7 & 1.30 & $4.4 \pm 0.6$ \\
uncover-v4\_prism-clear\_2561\_23955$^{\textcolor{blue}{L}}$ & 10.89 & 2.64 & 3.43 & 31.1 & 2.46 & $15.8 \pm 3.7$ \\
mom-uds01-v4\_prism-clear\_5224\_168214 & 11.46 & 2.60 & 3.08 & 53.9 & 3.50 & $10.4 \pm 1.7$ \\
mom-cos03-v4\_prism-clear\_5224\_247057 & 10.93 & 1.49 & 4.39 & 26.9 & 0.74 & $6.2 \pm 1.1$ \\
\bottomrule
\addlinespace
\multicolumn{7}{p{0.95\textwidth}}{Derived physical parameters for the golden sample, i.e., the sub-sample of galaxies with robust Na~I~D detection. $M_\star$ denotes the stellar mass in units of $M_\odot$, SFR is the star formation rate ($M_\odot\,\mathrm{yr}^{-1}$), $z_{\text{fit}}$ is the best-fit redshift, S/N50 refers to the median signal-to-noise ratio of the continuum, $A_V$ is the V-band dust attenuation, and EW represents the equivalent width of the Na I D absorption feature in $\AA$.

$^{\textcolor{blue}{a}}$\citealt{Glazebrook+17};
$^{\textcolor{blue}{b}}$\citealt{Setton+25};
$^{\textcolor{blue}{c}}$\citealt{Zhang+26}.
$^*$Broad-line AGN.
$^{\textcolor{blue}{L}}$Stellar mass corrected for gravitational lensing magnification.
}
\end{tabular}
\end{table*}

\section{Results}
\label{sec: results}

\subsection{Star Formation Main Sequence}
\label{sec: sfms}
Figure \ref{fig: sfms} illustrates the distribution of the 811 sources on the SFR-$M_*$ plane divided into 2 redshift bins, from $z=3$ to $z=3.5$, and $z>3.5$, selected to contain a comparable number of galaxies each.
We derive the recent SFR for each galaxy by averaging the SFH inferred by \texttt{prospector} over the three most recent time bins, corresponding to the most recent 100 Myr.
We compare our measurements with the Star-Forming Main Sequence (SFMS) relation defined by \cite{Simmonds_sfms+25} at the corresponding redshifts computed in the last 100 Myr.
The sample covers a broad range in stellar mass, extending from $\log(M_*/M_\odot) \sim 8$ to massive systems with $\log(M_*/M_\odot) > 11$; however, the galaxies with robust detections of Na~I~D tend to have large stellar masses.

To classify the evolutionary state of our galaxies, we define starburst galaxies as those lying more than $4\times$ above the SFMS relation, and quiescent (or quenched) galaxies as those lying more than $4\times$ below it. Based on these criteria, the majority of our sample follows the Main Sequence; however, we identify a significant population of outliers: 183 galaxies are classified as starbursts, while 62 galaxies are classified as quiescent/quenching, lying significantly below the relation.
Notably, with the adopted definition we find that 7 of the 20 galaxies ($\sim 35$\%) with robust detections are classified as quenched systems, while in the parent sample only 8\% of the galaxies are quiescent.

\begin{figure}[h] 
\centering
    \includegraphics[width=\linewidth]{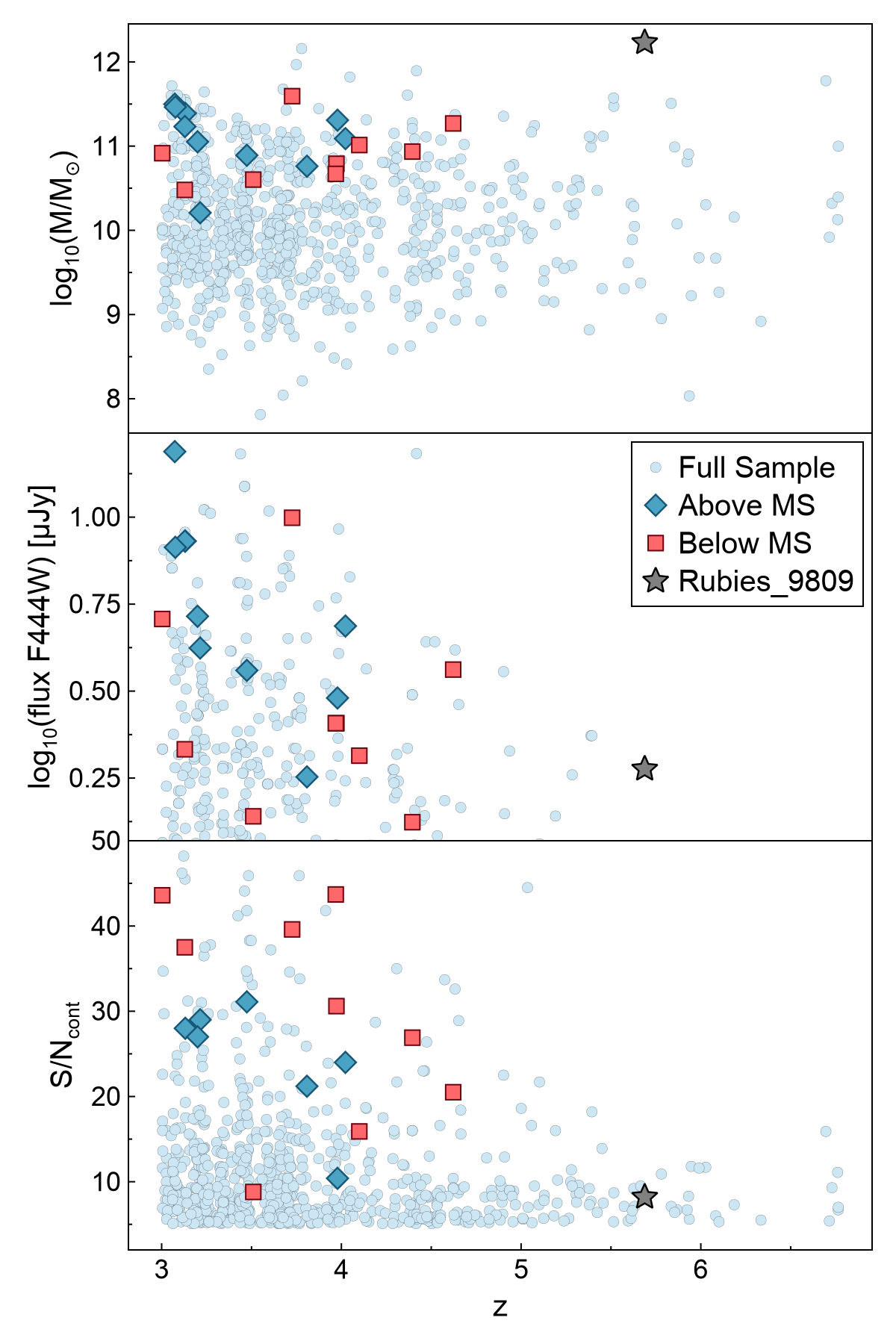}
    \caption{Galaxy properties for the parent sample, shown as a function of redshift. The vertical panels display, from top to bottom: stellar mass, flux density in the F444W filter, and the signal-to-noise ratio of the continuum ($S/N_{\mathrm{cont}}$). Square symbols identify galaxies in the golden sample (i.e., with robust Na~I~D detection), red squares mark systems below the Main Sequence, while blue squares mark systems on or above the Main Sequence.}
    \label{fig: sample prop}
\end{figure}

\begin{figure}[h]
\centering
\includegraphics[width=\linewidth]{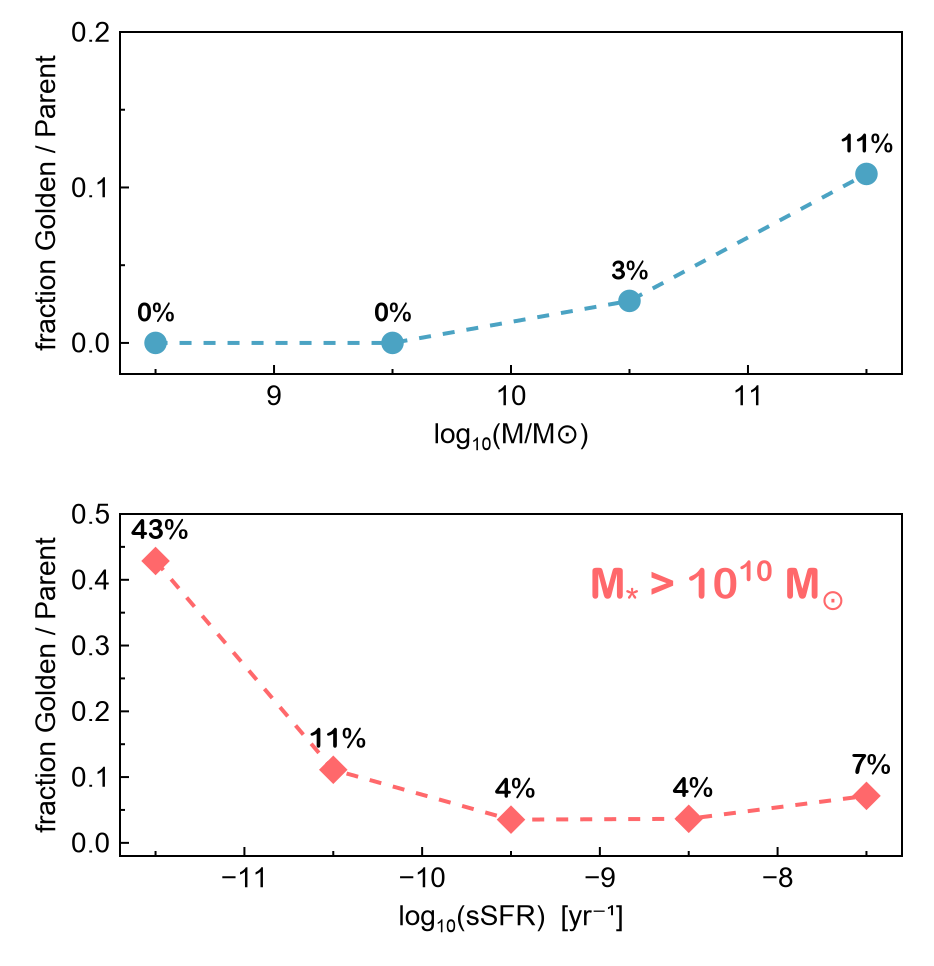}
\caption{Fraction of golden-sample galaxies (i.e., with robust Na~I~D detection) relative to the parent sample, in bins of stellar mass (\textbf{top}) and specific SFR (\textbf{bottom}), the latter restricted to galaxies with $M_\star > 10^{10}\,M_\odot$.}
\label{fig: golden fraction}
\end{figure}

\begin{figure}[p]
\centering
    \includegraphics[width=\linewidth]{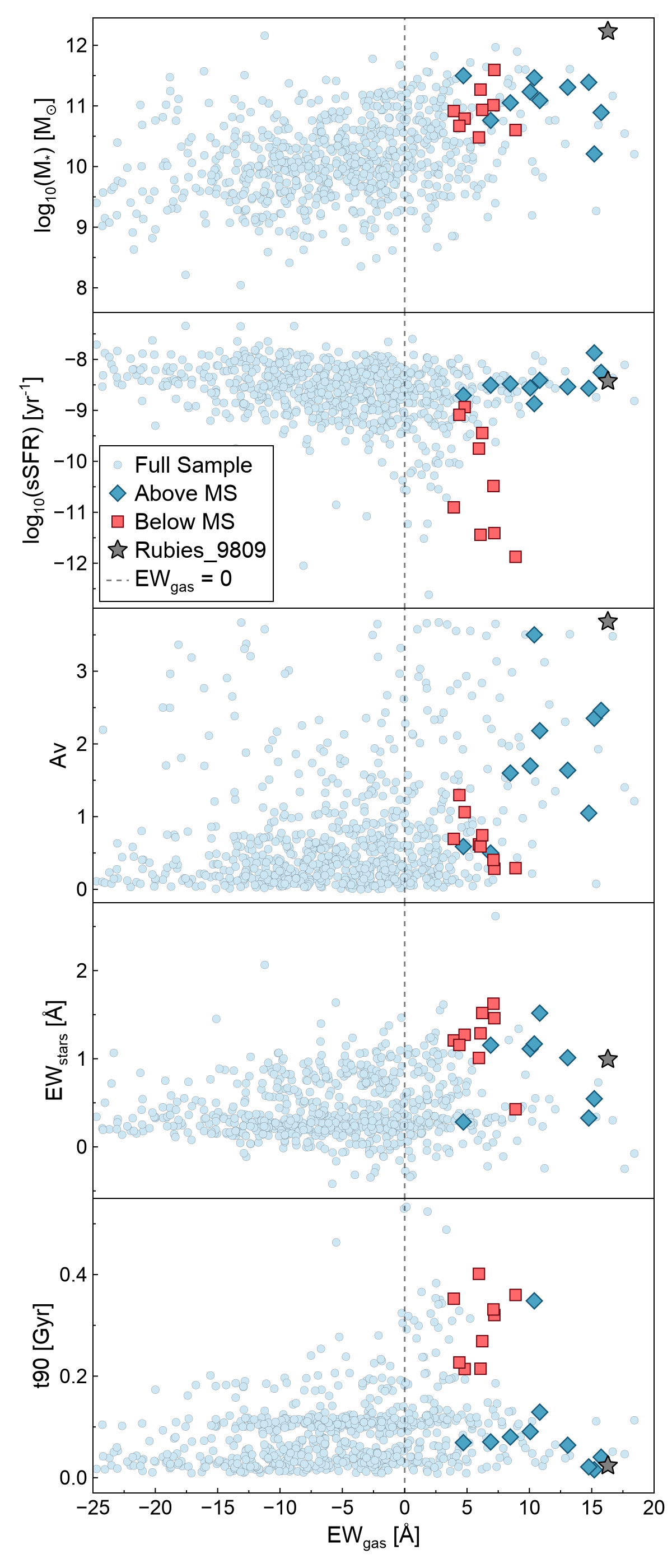}
    \caption{Galaxy properties in the parent sample as a function of EW$_{\mathrm{gas}}$, i.e., the excess Na I D equivalent width due to gas absorption (positive EW values) or emission (negative EW values). The panels display, from top to bottom: stellar mass, specific star formation rate ($\log \mathrm{sSFR}$), dust attenuation ($A_V$), the EW expected from the stellar models ($EW_{\mathrm{stars}}$), and assembly time ($t_{90}$). Square symbols identify galaxies in the golden sample (i.e., with robust Na~I~D detection), red squares mark systems below the Main Sequence, while blue squares mark systems on or above the Main Sequence.}
    \label{fig: ew mass av}
\end{figure}

\subsection{Host Galaxies Properties}
\label{sec: host gal prop}
The top panel of Figure \ref{fig: sample prop} shows the distribution of stellar mass and redshift for the parent sample. The golden sample of robust Na~I~D detections is also shown, split into the subsamples of galaxies that are above and below the Main Sequence. The Na~I detections are found in the lower half of the redshift range, $3 < z < 5$, except for a single galaxy at $z=5.7$, which is the only broad-line AGN. Also, the Na~I detections are concentrated in massive galaxies with $\log_{10}(M_*/M_{\odot}) > 10$. The mass trend is consistent with the result of recent JWST works \citep[e.g.,][]{davies_jwst_2024, zhu2026againneutraloutflowsz35, Taylor+26, Sun+26sample}.
To test whether this trend is simply due to the fact that it is easier to detect Na~I~D absorption in massive galaxies because they are brighter, in the bottom panels of Figure \ref{fig: sample prop} we show the distribution of F444W flux and SNR of the spectral continuum. Clearly, Na~I absorption is detected also in galaxies in the lower range of flux and SNR values, suggesting that the trend with stellar mass is real.

To quantify this trend, we compute the fraction of golden-sample galaxies relative to the parent sample in bins of stellar mass and specific SFR (sSFR), shown in Figure~\ref{fig: golden fraction}. The golden-sample fraction increases steadily with stellar mass, reaching $\sim11\%$ in the highest mass bin ($11 < \log(M_\star/M_\odot) < 12$). An even stronger trend is found with sSFR for massive ($M_\star > 10^{10} M_\odot$) galaxies: the detection fraction reaches $\sim43\%$ for galaxies with $\mathrm{sSFR} < 10^{-11}\,\mathrm{yr^{-1}}$, and remains substantial ($\sim11\%$) in the adjacent bin $10^{-11} < \mathrm{sSFR} < 10^{-10}\,\mathrm{yr^{-1}}$.

Figure~\ref{fig: ew mass av} shows the distribution of galaxy properties as a function of the EW measured from Na~I~D (after removal of the stellar template). The EW is positive when the line is detected in absorption, which indicates the presence of neutral gas. Due to the large uncertainties on the EW, many of the positive values do not represent statistically significant detections; for this reason we also show the golden sample of robust detections. Most of the parent sample, on the other hand, has negative EW values, indicating the presence of an emission line. Since Na~I~D is a resonant line, it is possible to observe it in emission, particularly in low-mass star-forming galaxies; however, in these systems the nearby He~I $\lambda 5876$ emission line is substantially stronger \citep{Concas+19, Borsani_&_Saintonge+19}. Since He~I and Na~I~D are blended in the low-resolution NIRSpec prism data, the negative EW values that we measure are likely due to the presence of strong emission by He~I.

We briefly describe the trend between Na~I EW and the physical properties shown in the panels of Figure~\ref{fig: ew mass av}, from top to bottom:
\begin{itemize}
    \item Stellar mass: as noted above, the Na~I detections are predominantly found in massive systems; however there is no trend between the amount of absorption and the galaxy stellar mass.

     \item Star formation activity: the sample with robust detections spans a wide range in sSFR. Many detections from the golden sample lie on the Main Sequence, while a non-negligible subset falls in quenched or quenching systems, suggesting that strong Na I D absorption is not limited to the most actively star-forming galaxies. However, the highest values of EW (EW$>10~\AA$) are found only above the Main Sequence, while more moderate values of EW are found both above and below the Main Sequence.
    
    \item Dust obscuration: The Na~I~D EW increases with $A_V$, consistent with previous studies at low and high redshift \citep[e.g.,][]{Poznanski+12, davies_jwst_2024}. This trend becomes even more pronounced when restricting to galaxies above the SFMS, where the highest EW values are found in the most dust-reddened systems. Galaxies below the SFMS with robust detections show correspondingly lower $A_V$ and lower EW values. This behavior is naturally explained by the fact that sodium remains neutral only when the gas is sufficiently shielded from ionizing radiation, making dust content a key prerequisite for the detection of Na~I~D absorption. 

    \item Stellar contribution: the fourth panel of Figure~\ref{fig: ew mass av} shows the EW of the Na~I absorption line produced in the stellar photospheres. We calculate this for each galaxy by fitting a Gaussian profile in absorption to the best-fit stellar model obtained by \texttt{prospector}. The stellar EW values span a range from 0 to 2 $\AA$, and are much lower than the EW values due to gas absorption for galaxies in the golden sample. We point out that the gas EW is measured after subtraction of the stellar template, so that it implicitly includes a correction for the stellar absorption. We thus conclude that the strongest observed Na~I absorption lines cannot be explained by the stellar continuum alone, even when allowing for systematic uncertainties in the stellar models due to, e.g., unknown abundance patterns.
    
    \item Assembly history: the last panel of Figure~\ref{fig: ew mass av} shows the relation between Na I D EW and $t_{90}$, the lookback time at which a galaxy assembled 90\% of its stellar mass. Small $t_{90}$ values correspond to systems that assembled most of their stellar mass recently, while larger values mark older and more evolved galaxies. Na~I detections populate a range of assembly histories, but the strongest absorption is concentrated at low $t_{90}$ and high sSFR, indicating that the most prominent neutral-gas signatures are associated with the rapid assembly phase of massive galaxies.
\end{itemize}

\begin{figure}[tbh]
    \centering
    \includegraphics[width=0.98\linewidth]{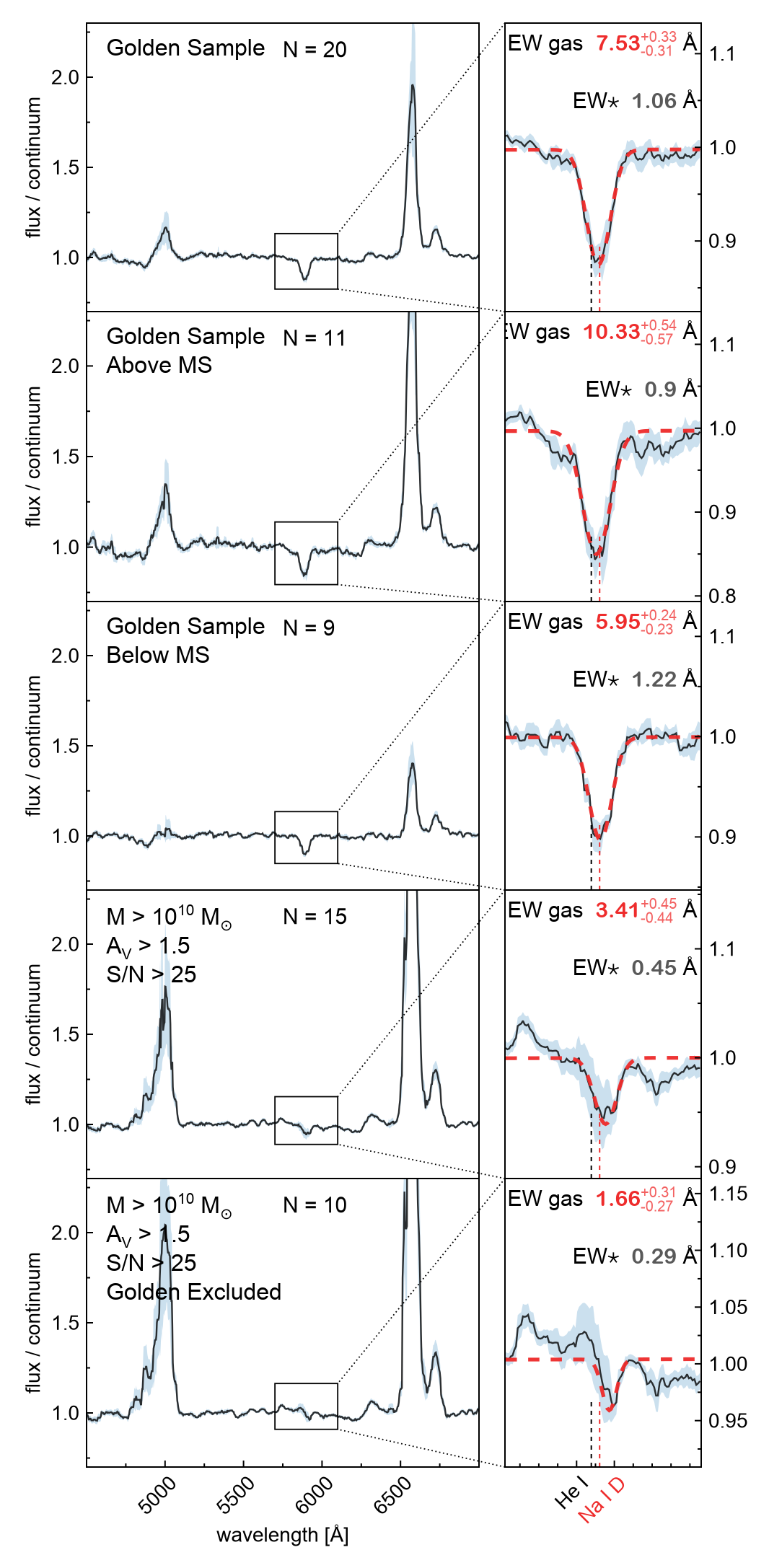}
    \caption{\textbf{Left:} Continuum-normalized mean stacks (black) with $1\sigma$ uncertainty (blue) for different subsamples: the golden sample (i.e., with robust Na~I~D detection), galaxies above and below the MS in the golden sample, high-extinction massive galaxies and high-extinction massive galaxies which are not part of the golden sample. \textbf{Right:} Zoom-in of the Na~I~D doublet for each of the stacks. Red dashed lines show the best-fit models. The $EW_{gas}$ measured from the continuum-normalized stack, and the mean stellar model $EW_{\star}$ are listed in each panel.
    Vertical lines mark the He~I and Na~I~D wavelengths.}    
    \label{fig: golden stack}
\end{figure}

\begin{figure*}[tbh] 
\centering
    \includegraphics[scale=0.675]{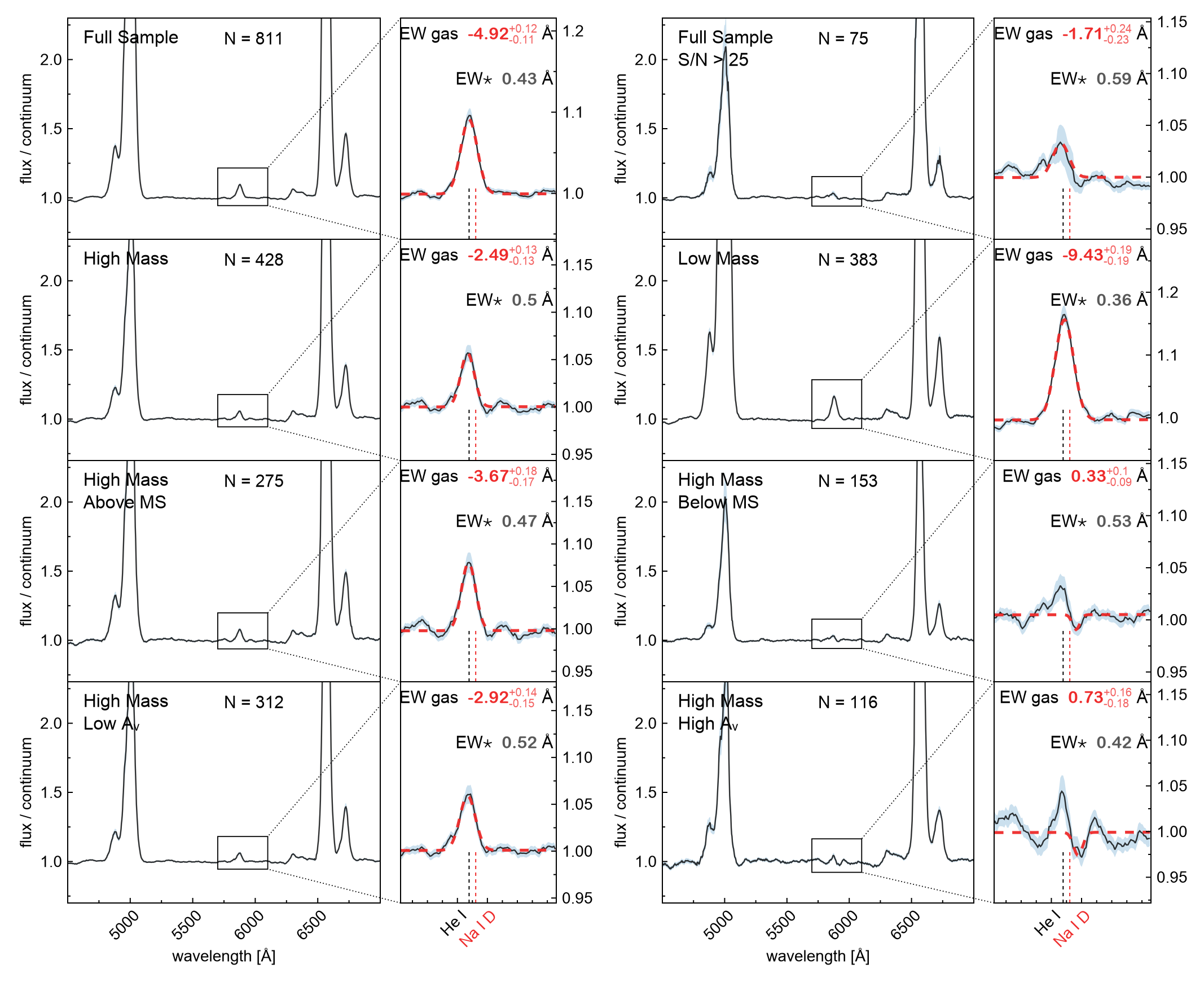}
    \caption{Stacked spectra for different subsamples of the parent galaxy sample. For each column, the left panels show the full rest-frame stacked spectra normalized by the stellar continuum, while the right panels present a zoom-in around the Na~I~D spectral region. The blue shaded regions indicate the $1\sigma$ uncertainty estimated through bootstrap resampling, and the dashed red line shows the best-fit model used to measure the Na~I~D equivalent width. The reported EW$_{\rm gas}$ values correspond to the gas component measured from the continuum-normalized stack (with negative values indicating gas emission), while EW$_\star$ indicates the average stellar contribution estimated from the \texttt{prospector} stellar models of the galaxies included in each stack. }
    \label{fig: stack}
\end{figure*}

\subsection{Stack Analysis}
To further investigate the Na~I~D absorption and its link to the galaxy physical conditions we construct a series of stacked spectra for different galaxy subsamples.
Prior to stacking, each individual spectrum is normalized by its corresponding best-fit stellar model obtained from \texttt{prospector}. The normalized spectra are then interpolated onto a common rest-frame wavelength grid using the algorithm presented in \cite{spectres}, minimizing the artificial super-sampling effects that can introduce additional uncertainties, and subsequently co-added using mean stacking combined with a 15$\sigma$ clipping procedure to reject strong outlier pixels. The contribution of each spectrum is weighted according to its continuum signal-to-noise ratio measured between rest-frame 4500 and 7000 $\AA$. We also perform an unweighted stack and confirm that the results do not change. The uncertainties associated with each pixel of the final stacked spectra are estimated through bootstrap resampling over 1000 iterations.
We then measure the EW of the Na~I~D absorption feature for each resulting stacked spectrum. The measurement is performed by fitting the spectral region with a model composed by a gaussian plus linear local continuum. The model parameters are optimized using a Markov Chain Monte Carlo (MCMC) approach employing the No-U-Turn Sampler (NUTS; \citealt{nuts}) as implemented in the Turing.jl library \citep{Turing.jl}. We define the lower and upper uncertainties as the 16th and 84th percentiles of the posterior distribution. Additionally, for each stack, we compute the stellar EW contribution as the mean of the individual $EW_{\rm stars}$ values derived from the \texttt{prospector} stellar models of the galaxies included in the stack.

Figure~\ref{fig: golden stack} shows the stacked spectra for the subsample of robust detections. The stack of the full golden sample composed of 20 galaxies is shown in the top panel, and the stacks of galaxies located above and below the SFMS are shown in the next two panels. All stacks exhibit a clear Na~I absorption feature. The strongest absorption is observed in galaxies above the SFMS, with an EW of $\sim10\,\AA$, consistent with the trends presented in Figure~\ref{fig: sfms}, where star-forming systems generally display larger Na~I equivalent widths. At the same time, a significant absorption feature with an EW of $\sim6\,\AA$ is also detected in galaxies below the SFMS, indicating that strong Na~I~D absorption is not limited to highly star-forming systems.

The results presented in the previous section revealed that the most significant individual detections of Na~I~D absorption are predominantly found in massive galaxies, and that the absorption EW increases with $A_V$. Motivated by these results, we construct an additional stack by selecting galaxies from the parent sample exclusively on the basis of their physical properties, namely high stellar mass ($M_\star > 10^{10}\,M_\odot$), large dust attenuation ($A_V > 1.5$), and high continuum Signal to noise ($\mathrm{S/N} > 25$), independently of whether they are classified as robust detections. The resulting stack contains 15 galaxies, of which 10 are not part of the golden sample. Despite this, the stacked spectrum reveals a clear Na~I~D absorption feature with an EW of $3.4\,\AA$. To verify that this signal is not solely driven by the golden-sample galaxies included in this stack, we also construct a stack restricted to the 10 galaxies in this selection that are not part of the golden sample. As shown in the last panel of Figure~\ref{fig: golden stack}, an absorption feature (EW $\sim1.7\,\AA$) is still detected in this reduced, purely mass, dust and S/N-selected subsample.
This result suggests that the golden sample likely represents only the most significant individual detections, while weaker Na~I~D absorption features are present in the broader population of galaxies with similar physical properties. The detection of an absorption profile in this stack further supports the idea that high stellar mass and large dust attenuation are key ingredients for the detection of neutral gas. 

To explore further correlations between neutral gas absorption and galaxy physical properties we also construct stacked spectra for different subsamples of the parent sample. Figure~\ref{fig: stack} presents the resulting stacks, including systems with a high S/N, massive ($M_\star > 10^{10}\,M_\odot$) and non-massive galaxies, systems above and below the SFMS, and galaxies with high and low dust attenuation ($A_V > 1.5$ and $A_V < 1.5$). Overall, none of the stacks extracted solely on the basis of these physical properties reveals unambiguous absorption.
The lack of a prominent absorption feature is likely related to the overshining effect produced by neighboring nebular emission lines, particularly He~I $\lambda5876$, which is blended with Na~I~D at the spectral resolution of NIRSpec prism observations. This effect is especially evident in the stack of the entire parent sample, where the integrated profile is dominated by emission. 
An intriguing result emerges when comparing the stacks of massive galaxies located above and below the SFMS. The stacked spectrum for massive galaxies below the MS shows marginal evidence of Na~I~D absorption, which may appear counterintuitive, since the strongest individual Na~I~D absorptions are typically found in star-forming galaxies, as noted above. However, this behavior can be explained by the lower intensity of nebular emission lines in less actively star-forming systems: the weaker He~I emission produces less contamination around the Na~I~D doublet, making the absorption feature easier to detect in the stacked spectrum.

\begin{figure}[h]
\centering

\includegraphics[width=\linewidth]{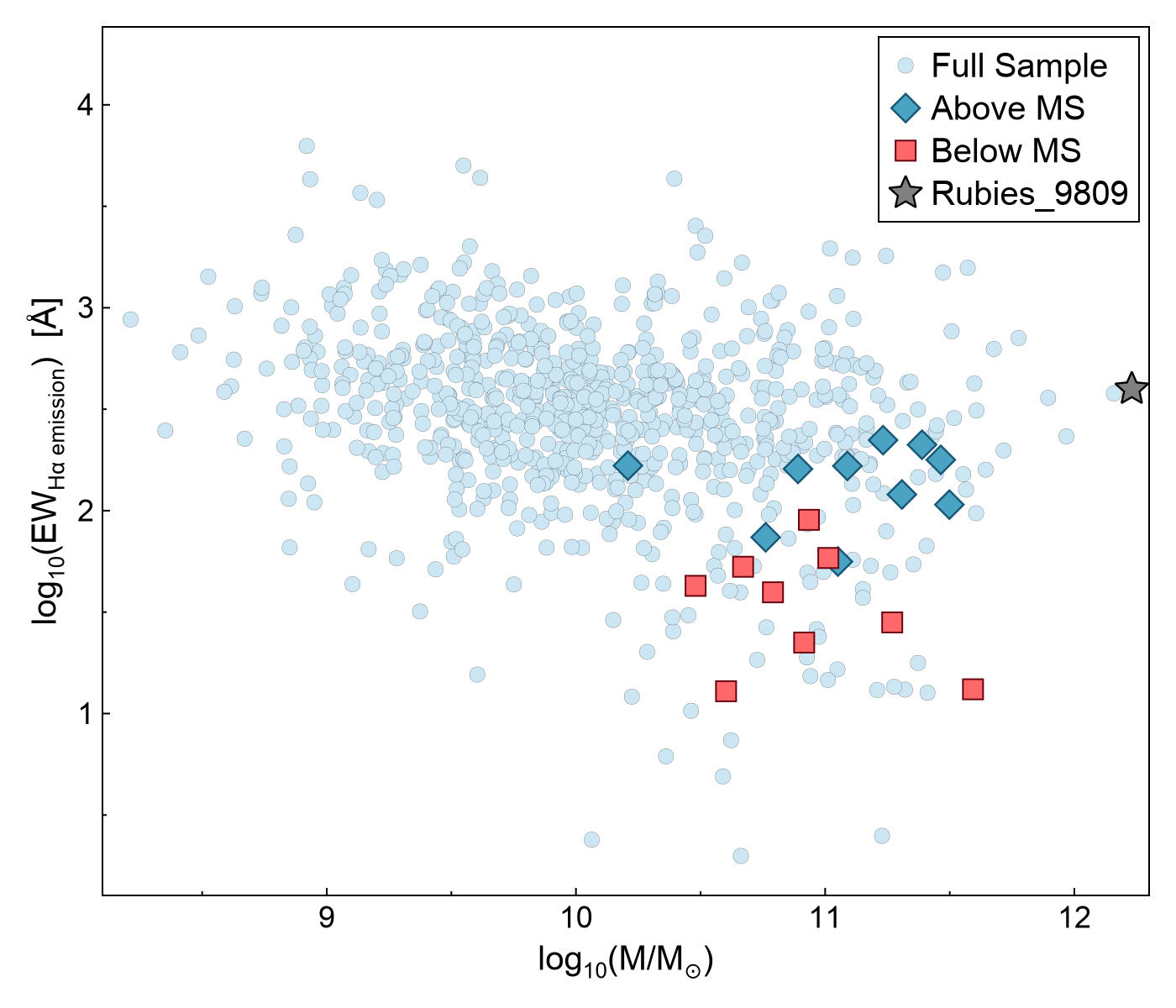}
\caption{H$\alpha$ EW in emission as a function of stellar mass.  Square symbols identify galaxies in the golden sample (i.e., with robust Na~I~D detection), red squares mark systems below the Main Sequence, while blue squares mark systems on or above the Main Sequence.}
\label{fig: na vs ha ew}
\end{figure}

To further test this interpretation, we examine how the Na~I~D detections depend on the strength of nearby nebular emission. Since He~I $\lambda5876$ cannot be measured directly at the resolution of our NIRSpec prism spectra, we use the H$\alpha$ equivalent width as a proxy for the strength of the nebular emission responsible for contaminating the Na~I~D region. We find that Na~I~D detections preferentially occur in galaxies with lower H$\alpha$ EW, consistent with He~I emission being one of the main factors limiting our ability to detect Na~I~D absorption. Figure~\ref{fig: na vs ha ew} illustrates this trend.

Our results indicate that neutral gas absorption is common in massive galaxies at $z>3$, and particularly in quiescent and dusty star-forming systems. However, at the spectral resolution of the NIRSpec prism, Na~I~D absorption can be detected only when strong nebular emission lines are not present, which partly explains the higher incidence of detections in quiescent and dusty galaxies. We thus expect the true incidence of neutral gas absorption to be substantially higher than what is measured in the full sample.

\begin{table*}[thb]
\centering
\caption{Estimated neutral outflow properties derived from the stacked spectra of the golden sample.}
\label{tab:outflow_properties}
\begin{tabular}{lccccc}
\hline
Sample & EW Na~I~D [$\AA$] & $N_{\rm Na\,I}$ [$\mathrm{cm^{-2}}$] & $N_{\rm H}$ [$\mathrm{cm^{-2}}$] & $M_{\rm out}$ [$M_\odot$] & $\dot{M}_{\rm out}$ [$M_\odot\,\mathrm{yr^{-1}}$] \\
\hline
golden sample                & $7.5$  & $2.4 \times 10^{13}$ & $7.7 \times 10^{20}$ & $4.4 \times 10^{7}$ & $8.9$ \\
golden sample (Below SFMS)   & $5.9$  & $1.9 \times 10^{13}$ & $6.1 \times 10^{20}$ & $3.4 \times 10^{7}$ & $7.1$ \\
golden sample (Above SFMS)   & $10.3$ & $3.4 \times 10^{13}$ & $1.1 \times 10^{21}$ & $6.0 \times 10^{7}$ & $12$  \\
Massive + Dusty + High S/N   & $3.4$  & $1.1 \times 10^{13}$ & $3.5 \times 10^{20}$ & $2.0 \times 10^{7}$ & $4.0$ \\
Massive + Dusty + High S/N (excluding Golden) & $1.7$ & $5.4 \times 10^{12}$ & $1.7 \times 10^{20}$ & $9.6 \times 10^{6}$ & $2.0$ \\
\hline
\end{tabular}
\end{table*}


\section{Outflow Properties}
\label{sec: outflows}

\subsection{EW Excess as an Outflow Tracer} \label{sec: Na excess}
\begin{figure}[tb] 
\centering
    \includegraphics[width=\linewidth]{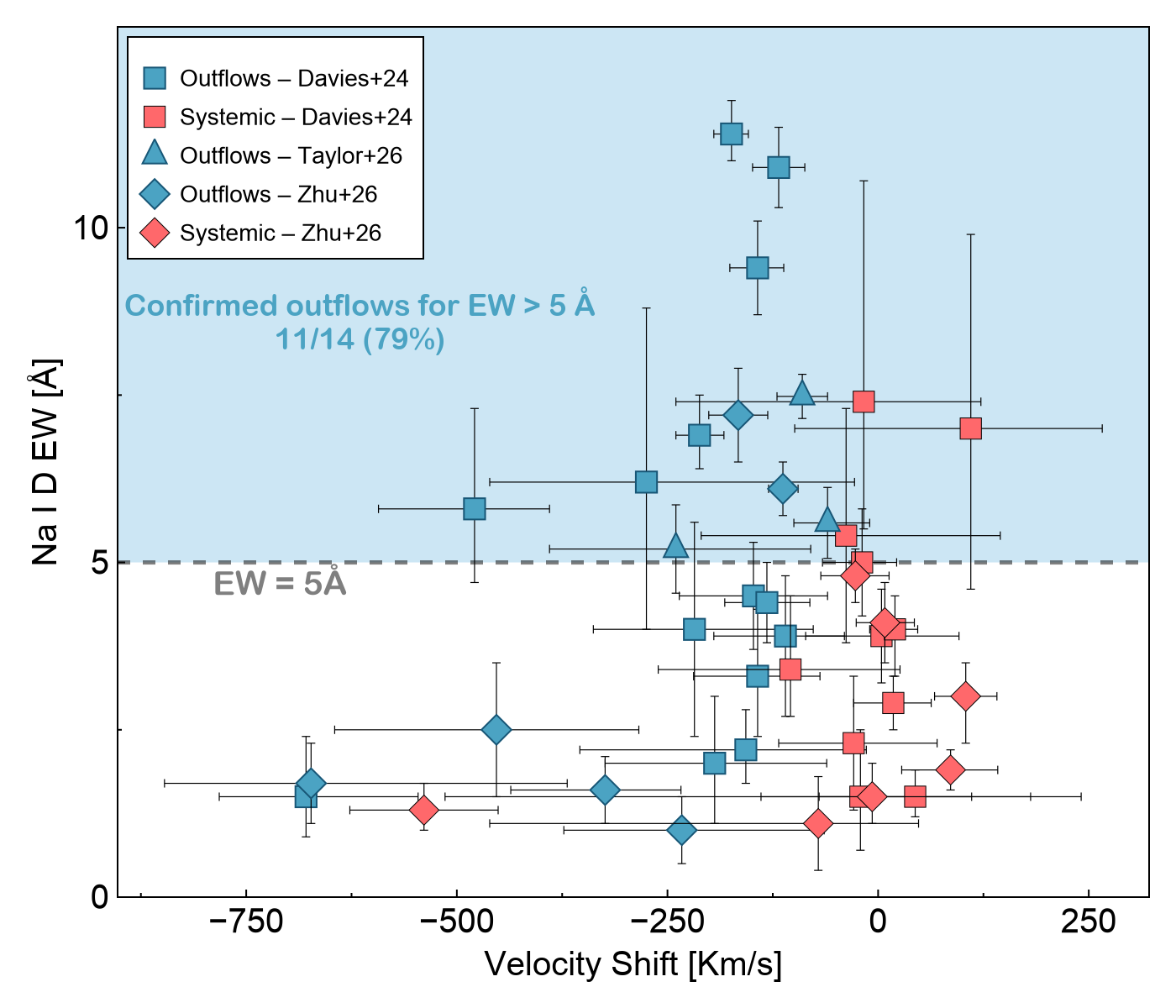}
    \caption{Literature compilation of Na~I~D EW versus velocity shift for galaxies at $z \gtrsim 2$ observed with medium-resolution NIRSpec observations \citep{davies_jwst_2024, Taylor+26, zhu2026againneutraloutflowsz35}. Blue symbols represent systems with blueshifted Na~I, which are therefore classified as outflows, while red symbols represent systemic components associated with the galaxy rest frame. Different marker shapes identify measurements from different literature samples. The horizontal dashed line indicates a reference equivalent-width threshold of $5\,\AA$.}
    \label{fig: ew davies 2024}
\end{figure}

Although the low spectral resolution of our prism observations precludes a detailed kinematic analysis, recent JWST studies have shown that large Na~I~D EWs are frequently associated with blueshifted absorption profiles, which are clear signatures of outflowing neutral gas \citep[e.g.,][]{davies_jwst_2024, Taylor+26, zhu2026againneutraloutflowsz35}. To investigate this connection, we compile published measurements of Na~I~D EW and velocity shift measured from medium-resolution NIRSpec observations, which allow a robust characterization of the neutral gas kinematics. In particular, we consider galaxies at $z \gtrsim 2$ from the Blue Jay \citep{BlueJay-Survey}, EXCELS \citep{Taylor+26}, and DeepDive \citep{zhu2026againneutraloutflowsz35} surveys.
Figure~\ref{fig: ew davies 2024} shows the distribution of EW and velocity shift for these samples, highlighting the galaxies with blueshifted Na~I~D absorption. The strongest absorption systems are predominantly associated with blueshifted profiles, suggesting that large Na~I~D EWs can be used as a proxy for powerful neutral outflows. Adopting $5\,\AA$ as an approximate threshold for strong absorption, we find that within the literature compilation 11 out of the 14 galaxies (79\%) above this threshold show blueshifted Na~I~D absorption. This is a lower limit on the incidence of outflows among the galaxies with strong Na~I~D absorption, because some of the systemic velocity measurements may be due to low outflow velocities or inclination effects, rather than to the absence of an outflow. We therefore conclude that the galaxies entering our golden sample, which have EW values mostly above the $5\,\AA$ threshold, are likely tracing neutral outflows. Given the wide diversity in the galaxy properties, it is possible that both star-formation-driven and AGN-driven outflows are present in our sample \citep[see also][]{davies_jwst_2024, Sun+26sample}.

\subsection{Incidence and Opening Angle}\label{sec: opening angle}

The fraction of galaxies in which a Na~I~D outflow is detected places a lower limit on the covering fraction of the outflowing gas. We restrict this analysis to massive galaxies ($M_\star > 10^{10}\,M_\odot$) and split the sample by star formation activity and dust content. In the star-forming population (on or above the SFMS) we robustly detect 11 outflows out of 275 massive galaxies, giving a detection fraction $f_\mathrm{det}^\mathrm{SF} = 4\%$. This is lower than similar medium- or high-resolution studies at $z\sim2$ \citep[e.g.][]{davies_jwst_2024, Sun+26sample} likely because of He~I contamination. In the quenched population we detect 7 outflows out of 53 massive galaxies, giving $f_\mathrm{det}^\mathrm{QU} = 13.2\%$. Among massive galaxies with high dust attenuation ($A_V > 1.5$) we detect 8 outflows out of 116 galaxies, giving $f_\mathrm{det}^{A_V} = 6.9\%$. The incidence is highest among quenched galaxies, followed by dusty, high-$A_V$ systems.
The detection fractions translate into geometric constraints on the outflow solid angle. Under the assumption of randomly oriented galaxies, an absorption-selected survey detects an outflow only when the line of sight passes through the outflow cone. Consequently, the observed detection fraction is a direct estimate of the solid-angle covering fraction:
\begin{equation}
    f_\mathrm{det} \equiv \frac{N_\mathrm{det}}{N_\mathrm{tot}} \leq f_\mathrm{cov} \equiv \frac{\Omega}{4\pi}\,
\end{equation}
where the inequality holds because outflows below our detection threshold or whose absorption is diluted by He~I $\lambda5876$ emission are missed, so the true incidence is necessarily higher than the observed one.

For a biconical outflow with half-opening angle $\theta$, the solid angle subtended by both cones is $\Omega = 2 \times 2\pi(1-\cos\theta) = 4\pi(1-\cos\theta)$, so:
\begin{equation}
    f_\mathrm{cov} = \frac{\Omega}{4\pi} = 1 - \cos\theta\,.
\end{equation}
Since $f_\mathrm{det} \leq f_\mathrm{cov}$, the observed detection fraction yields a lower limit on the half-opening angle:
\begin{equation}
    \theta \geq \arccos\!\left(1 - f_\mathrm{det}\right)\,.
\end{equation}
Applying this to the three populations we obtain $\theta^\mathrm{SF} \gtrsim 16^\circ$, $\theta^\mathrm{QU} \gtrsim 30^\circ$, and $\theta^{A_V} \gtrsim 21^\circ$.

In practice, the true opening angles are likely larger than these lower limits, since our detection efficiency is reduced by sensitivity limitations and He~I contamination, which preferentially suppress detections in star-forming systems. The higher incidence observed in dusty and quenched galaxies, where He~I emission is weaker, is thus at least partly due to our improved sensitivity in those systems, but may also reflect a genuinely higher incidence, or a wider opening angle in systems where the outflow is no longer collimated by dense star-forming gas.

\subsection{Outflow Mass Estimates}
We adopt a simplified approach to estimate the mass of the outflowing neutral gas, based on a set of conservative assumptions motivated by the observational constraints discussed above.
We first calculate the Na~I column density from the measured equivalent width assuming that the gas is optically thin \citep{Draine+2011_book}:
\begin{equation}
    N_{\rm Na\,I} =
    1.13 \times 10^{20}
    \frac{{\rm EW}}{{f_{lu}\lambda_{lu}^2}}
    \quad {\rm cm}^{-2},
\end{equation}
where EW and $\lambda_{lu}$ are expressed in \AA, and $f_{lu}$ is the oscillator strength. Na~I~D is a doublet, but we cannot resolve the two components because of the low spectral resolution. We thus measure the total EW, take $\lambda_{lu}$ to be the mean wavelength of the doublet, and adopt $f_{lu} \approx 1$ which is the sum of the oscillator strengths for the two transitions in the doublet.

The hydrogen column density is then inferred assuming the empirical relation from \citealt{Moretti+26}:
\begin{equation}
    \log N_{\rm H} = \log N_{\rm Na\,I} + 7.5.
\end{equation}

Assuming a spherical geometry, the outflow mass can be written as \citep{Rupke+05}:
\begin{equation}
    M_{\rm out} = 1.4\,m_p\,\Omega\,N_{\rm H}\,R_{\rm out}^2,
\end{equation}
where the factor 1.4 accounts for the mass in helium, $m_p$ is the proton mass, $\Omega$ is the solid angle subtended by the outflow, and $R_{\rm out}$ is the characteristic outflow radius. We adopt a conservative outflow radius of $R_{\rm out}=1\,{\rm kpc}$ and a covering geometry corresponding to $\Omega/4\pi = 0.4$ \citep{belli_star_2024}, i.e., a half-opening angle of $\theta \approx 53^\circ$; this is consistent with, though larger than, the lower limits on the outflow opening angle derived in Sect.~\ref{sec: opening angle} above.

To estimate the characteristic outflow velocity, we use the literature compilation presented in Figure~\ref{fig: ew davies 2024}. The galaxies showing the strongest Na~I~D absorption, comparable to the systems in our golden sample, typically exhibit blueshifted velocities of a few hundred ${\rm km\,s^{-1}}$. Motivated by this distribution, we adopt a representative value of $v_{\rm out}=200\,{\rm km\,s^{-1}}$.
Finally, the mass outflow rate is estimated as
\begin{equation}
    \dot{M}_{\rm out} =
    \frac{M_{\rm out}}{R_{\rm out}}\,v_{\rm out}.
\end{equation}

If we combine all the previous steps we obtain a linear relation between the measured EW of Na~I~D and the mass outflow rate:
\begin{equation}
    \dot{M}_{\rm out} = 1.2~M_\odot~ \mathrm{yr}^{-1} \left( \frac{\Omega}{ 0.4 \cdot 4\pi} \right) \left( \frac{R_{\rm out}}{1~{\rm kpc}} \right) \left( \frac{v_{\rm out}}{200~{\rm km\,s^{-1}}} \right) \left( \frac{{\rm EW}}{1~\AA} \right)
\end{equation}

The inferred neutral outflow properties for the different stacked samples are summarized in Table~\ref{tab:outflow_properties}. The full golden sample yields a characteristic Na~I column density of the order of $\sim10^{13}\,{\rm cm}^{-2}$, corresponding to hydrogen column densities of $\sim10^{20}\,{\rm cm}^{-2}$, neutral outflow masses of a few $\times10^{7}\,M_\odot$, and mass outflow rates in the range $\sim4$--$12\,M_\odot\,{\rm yr}^{-1}$ depending on the considered subsample.

Among all stacks, galaxies located above the SFMS exhibit the largest inferred mass outflow rates, reaching $\dot{M}_{\rm out} \sim 12\,M_\odot\,{\rm yr}^{-1}$. Lower but still substantial values are obtained for galaxies below the SFMS, while the physically selected stack of massive, dusty galaxies with high continuum S/N also shows a significant neutral gas outflow rate of $\dot{M}_{\rm out} \sim 4\,M_\odot\,{\rm yr}^{-1}$.

These values should be interpreted as conservative estimates. In particular, the assumed outflow radius represents one of the dominant sources of uncertainty. We adopt $R_{\rm out}=1\,{\rm kpc}$ as a conservative value, but characteristic outflow sizes reported in the literature typically span the range $\sim1$--$3\,{\rm kpc}$ \citep{belli_star_2024, D_Eugenio+24}. Adopting a larger radius, for example $R_{\rm out}=3\,{\rm kpc}$, would increase the inferred outflow mass by nearly an order of magnitude and the mass outflow rate by a factor of three. Despite the conservative nature of our assumptions, the inferred $\dot{M}_{\rm out}$ is larger than the star formation rate for 6 out of 20 galaxies in the golden sample ($\approx30\%$).

These results support a picture in which neutral outflows may already play a non-negligible role in regulating the gas content of massive galaxies at $z>3$.

\section{Summary and Conclusions}
\label{sec: conclusions}

\subsection{Summary}
In this work, we carried out the first systematic search for strong neutral-gas absorption traced by Na~I~D in a large archival sample of 811 galaxies at $3<z<7$ observed with the NIRSpec prism. By combining low-resolution spectroscopy with detailed SED fitting performed using \texttt{prospector}, we isolated the non-stellar Na~I~D component and investigated its connection with the physical properties of high-redshift galaxies.

We defined a golden sample of 20 galaxies with S/N\,$>3$ on the Na~I~D equivalent width, where the signal-to-noise is computed on the continuum-subtracted spectral residuals using a rescaled error spectrum that accounts for the underestimated flux uncertainties. The strongest absorption features are preferentially associated with massive galaxies above the SFMS, with high dust attenuation and rapidly assembling stellar populations. At the same time, $\sim35\%$ of the golden sample consists of systems with suppressed star formation activity, suggesting that neutral outflows remain present even after the peak growth phase of the galaxy.

A key result of this work is that the observed Na~I~D equivalent widths cannot be reproduced by stellar population models alone. The stellar contribution inferred from the \texttt{prospector} fits typically remains below $\sim2 \AA$, while the observed gas absorption reaches values in the range $4$ to $16\AA$ after removal of the inferred stellar absorption. This constitutes strong evidence for the presence of substantial amounts of neutral gas in these systems. By comparing the measured EWs of Na~I~D to those of recent medium-resolution JWST studies, we conclude that a significant fraction of our detections likely trace outflowing neutral gas.

The stacking analysis further strengthens this picture. Significant Na~I~D absorption is recovered not only in the stack of the robust detections, but also in stacks selected purely on the basis of physical properties such as high stellar mass and large dust attenuation. This suggests that the golden sample represents the most extreme tail of the broader population of galaxies hosting neutral absorption features that remain individually undetected in NIRSpec prism.

Using simple, conservative assumptions, we estimate neutral outflow masses of a few $\times10^7\,M_\odot$ and mass outflow rates of $\sim4$--$12\,M_\odot\,\mathrm{yr}^{-1}$ in the stacked spectra. The inferred outflow rate is comparable to or larger than the ongoing star formation rate for $\sim30\%$ of galaxies in the golden sample, indicating that neutral outflows play an important role in regulating the gas content.

The incidence of neutral gas detection depends on how the sample is split. Restricting to massive galaxies formally classified as below the SFMS, the detection fraction is $f_\mathrm{det}^\mathrm{QU} = 13\%$, implying an outflow opening angle $\theta^\mathrm{QU} \gtrsim 30^\circ$, higher than in the dusty population ($A_V>1.5$, $f_\mathrm{det}^{A_V} = 6.9\%$, $\theta^{A_V} \gtrsim 21^\circ$). An even more striking trend emerges when galaxies are instead selected by an absolute sSFR threshold: among galaxies with $\mathrm{sSFR}<10^{-11}\,\mathrm{yr^{-1}}$, the golden-sample fraction reaches $\sim43\%$. This trend, however, is likely shaped at least in part by a selection effect intrinsic to the prism resolution: as He~I $\lambda5876$ is blended with Na~I~D, strong nebular emission can fill in and mask the absorption signal, reducing our sensitivity in actively star-forming systems. Using H$\alpha$ EW as a proxy for the strength of this contaminating emission, we find that across the full sample Na~I~D detections are concentrated at low H$\alpha$ emission EW, regardless of star formation activity. We therefore cannot firmly establish whether the incidence of neutral outflows is intrinsically higher in quiescent and dusty systems, or whether our sensitivity to Na~I~D absorption is simply enhanced in these populations due to their weaker nebular emission.

\subsection{Conclusions}
Our results suggest that neutral outflows may already be present in massive galaxies at $z>3$, during the epoch when the first quiescent systems emerge. Two galaxy populations appear to host the strongest Na~I~D features. The first are dusty star-forming galaxies, which are likely undergoing intense obscured star formation. These systems may represent the immediate progenitors of the compact quiescent galaxies observed at similar and lower redshifts, with the powerful outflows they drive potentially contributing to the expulsion or heating of the remaining gas reservoir and thus to the suppression of star formation. The second population consists of already-quiescent or rapidly quenching systems, in which significant neutral absorption persists even after star formation has been largely suppressed. The presence of strong neutral outflows in both dusty star-forming and quiescent galaxies is consistent with a scenario in which the feedback mechanism plays an important role in the transition between these two populations.
The detection fraction is significantly higher in quiescent galaxies, reaching $\sim43\%$ among the most massive quenched systems. This enhancement may partly reflect an observational bias rather than a purely intrinsic trend: in quiescent galaxies the nebular emission lines are intrinsically weak, and in particular the He~I $\lambda5876$ line that blends with and fills in the Na~I~D absorption at the prism resolution is much fainter than in actively star-forming systems. Our sensitivity to Na~I~D absorption is therefore enhanced in the absence of strong He~I emission, which may boost the apparent incidence of neutral gas in quenched galaxies.
An important limitation of our analysis concerns the detection of Na~I~D absorption in dust-poor, highly star-forming galaxies. In such systems, the He~I $\lambda$5876 emission line can partially or entirely fill in the absorption signal, reducing our sensitivity to neutral gas in this regime. We therefore cannot establish whether dust-free star-forming galaxies at $z>3$ also host neutral outflows. However, we note that this is unlikely to be a purely observational bias: the survival of neutral sodium requires effective shielding from the UV radiation field, a condition most naturally met in dusty environments where gas and dust are co-spatial.
Our findings support a scenario in which cold-gas feedback is already a key process in the early Universe, operating across a range of galaxy evolutionary stages. While the low spectral resolution of the prism prevents a detailed kinematic characterization, a comparison with previously confirmed outflows with similar Na I D EW, together with the observed connection between neutral gas and dust-rich and quiescent systems, suggests significant neutral-gas-driven feedback at $z>3$.

Future medium- and high-resolution JWST observations of larger samples will be essential to directly resolve the kinematics of these systems and test the scenario outlined here. In particular, resolved observations will allow us to disentangle Na~I~D absorption from He~I emission in highly star-forming galaxies, directly probing whether dust-free systems at $z>3$ also drive neutral outflows, and to quantify the physical mechanisms responsible for launching them. Such studies will ultimately determine the relative contribution of neutral outflows to the quenching of star formation across the diverse galaxy population in the early Universe.

\begin{acknowledgements}
This research is supported by the ERC grant 101076080 ``Red Cardinal''. RW acknowledges funding of a Leibniz Junior Research Group (project number J131/2022). This research was supported by the International Space Science Institute (ISSI) in Bern, through ISSI International Team project 24-602 ``Multiphase Outflows in Galaxies at Cosmic Noon''. This work is based on observations made with the NASA/ESA/CSA James Webb Space Telescope. The data were obtained from the Mikulski Archive for Space Telescopes at the Space Telescope Science Institute, which is operated by the Association of Universities for Research in Astronomy, Inc., under NASA contract NAS 5-03127 for JWST. The data products presented herein were retrieved from the Dawn JWST Archive (DJA). DJA is an initiative of the Cosmic Dawn Center (DAWN), which is funded by the Danish National Research Foundation under grant DNRF140. P.D. warmly acknowledges support from an NSERC discovery grant (RGPIN-2025-06182). 
\end{acknowledgements}

\appendix

\section{AGN presence}\label{sec: app-agn}
Our sample selection criteria do not explicitly exclude the presence of AGNs. We therefore perform an analysis to quantify the potential AGN presence within our sample.

\begin{figure*}[tbh]
  \centering

  \begin{subfigure}{0.48\textwidth}
    \centering
    \includegraphics[width=\linewidth]{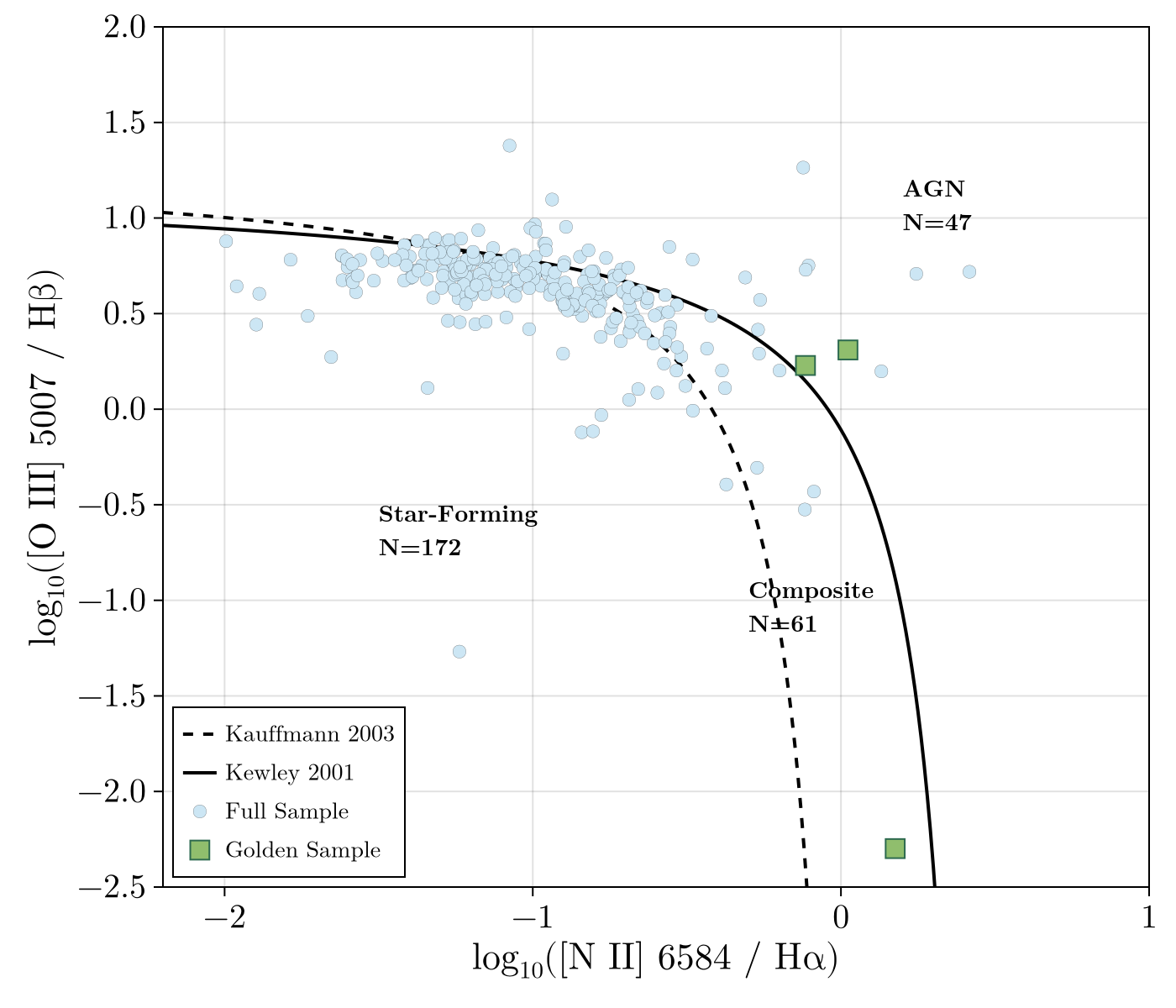}
    \caption{}
    \label{fig:bpt}
  \end{subfigure}
  \hfill
  \begin{subfigure}{0.48\textwidth}
    \centering
    \includegraphics[width=\linewidth]{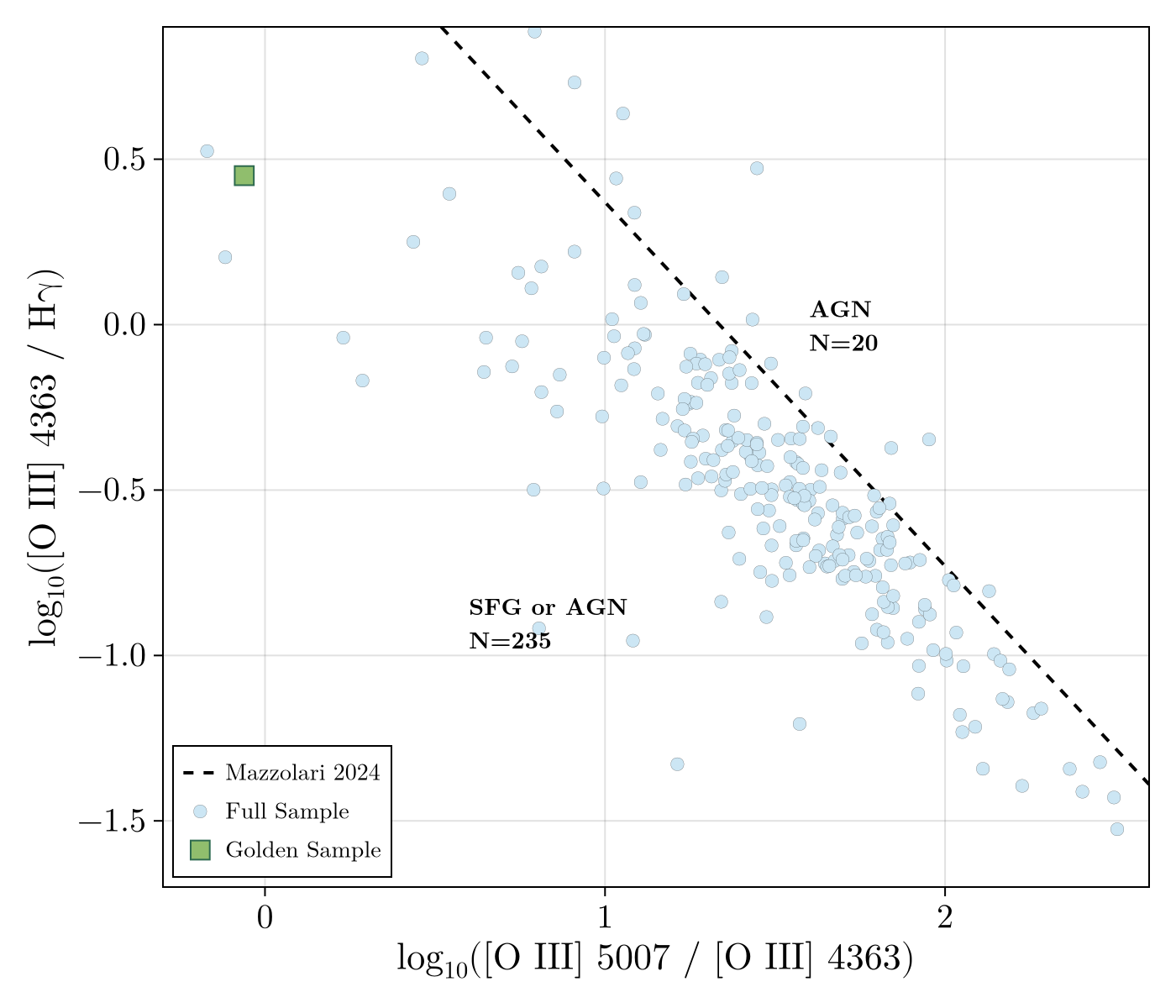}
    \caption{}
    \label{fig:mazzolari}
  \end{subfigure}

    \caption{\textbf{Panel a:} BPT diagnostic diagram, illustrating the flux ratio $\log_{10}([\text{O III}]\, 5007 / \text{H}\beta)$ versus $\log_{10}([\text{N II}]\, 6584 / \text{H}\alpha)$. The solid black line \citep{Kewley_agn} serves as the empirical upper limit for Star-Forming Galaxies, delineating the boundary from AGN. The dashed gray line \citep{Kauffmann_agn} separates SFGs from Composite sources. The sample is classified into Star-Forming ($N=172$), Composite ($N=61$), and AGN ($N=47$) sources. Green squares mark galaxies in the golden sample, i.e. with robust Na I D detection. \textbf{Panel b:} high-ionization diagnostic diagram from \citealt{Mazzolari_2024}, plotting $\log_{10}([\text{O III}]\, 4363 / \text{H}\gamma)$ against $\log_{10}([\text{O III}]\, 5007 / [\text{O III}]\, 4363)$. The dashed black line defines the boundary separating sources primarily powered by AGN activity ($N=20$) from those dominated by Star-Forming Galaxies or with ambiguous classifications ($N=235$).}

  \label{fig:gruppo}
\end{figure*}

This identification is challenging with our primary dataset, as the low resolution of the prism spectra blends the classic optical emission lines (e.g., H$\alpha$, $[NII]$, $H\beta$, $[OIII]$) typically used for AGN diagnostics. We therefore leverage external data, utilizing pre-computed emission line fluxes from the DJA catalog to construct diagnostic diagrams for the subset of our 811-galaxy sample for which medium- or high-resolution spectra are available. We successfully retrieve fluxes to construct the [OIII]$\lambda 5007$/H$\beta$ versus [NII]$\lambda 6584$/H$\alpha$ diagram (BPT; \citealt{BPT}) for 280 galaxies (Figure \ref{fig:bpt}). From the BPT analysis, 47 of the 280 objects are classified as potential AGNs, using the demarcation line defined by \cite{Kauffmann_agn} and \cite{Kewley_agn}. 

It is well known that the classical BPT diagram is not always a reliable AGN diagnostic at high redshift. Many high-$z$ JWST-detected galaxies show very high values of the ionization parameter, low metallicity, and high SFR and gas densities compared to local galaxies, so they tend to lie close to, or scatter across, the demarcation line traditionally used to separate star-forming galaxies from AGN in the BPT diagram \citep{Mazzolari_2024}. For this reason, in addition to the BPT diagram we also employ the high-ionization diagnostic diagram from \cite{Mazzolari_2024}, employing the [OIII]$\lambda 5007$/[OIII]$\lambda 4363$ versus [OIII]$\lambda 4363$/H$\gamma$ line ratios, which provides a more robust classification in this regime.
We retrieve fluxes for the  Mazzolari diagram for 255 galaxies (Figure \ref{fig:mazzolari}); 20 of these are classified as potential AGNs. We thus conclude that AGN contamination is not particularly strong in our sample.

\bibliographystyle{aa} 
\bibliography{bibliography}

\end{document}